\documentclass[aps,showpacs,preprintnumbers,amsmath,amssymb,nofootinbib,preprint,showkeys]{revtex4}

\usepackage{graphicx}
\usepackage{color}
      
\def \be  {\begin{equation}}
\def \ee  {\end{equation}}
\def \ee  {\end{equation}}
\def \bea {\begin{eqnarray}}
\def \eea {\end{eqnarray}}

\def \Tr  {\bf{Tr}}

\begin{document}

\preprint{ECTP-2012-01}

\title{On the Higher Moments of Particle Multiplicity, Chemical Freeze-Out and QCD Critical Endpoint}

\author{A.~Tawfik}
\email{a.tawfik@eng.mti.edu.eg}
\email{atawfik@cern.ch}
\affiliation{Egyptian Center for Theoretical Physics (ECTP), MTI University, Cairo, Egypt}
\affiliation{Research Center for Einstein Physics, Freie-University Berlin, Berlin, Germany}

\date{\today}

\begin{abstract}
We calculate the first six non-normalized moments of particle multiplicity within the framework of
the hadron resonance gas model. In terms of the lower order moments and corresponding
correlation functions, general expressions of higher order moments are derived. Thermal
evolution of the first four normalized moments and their products (ratios) are studied at different
chemical potentials μ, so that it is possible to evaluate them at chemical freeze out curve. It is
found that a non-monotonic behaviour, reflecting the dynamical fluctuation and strong
correlation of particles starts to appear from the normalized third order moment. We introduce
novel conditions for describing the chemical freeze out curve. Although the hadron resonance
gas model does not contain any information on the criticality related to the chiral dynamics and
singularity in the physical observables, we are able find out the location of the QCD critical
endpoint at $\mu\sim350~$ MeV and temperature $T \sim 162~$MeV.

\end{abstract}


\maketitle


\section{Introduction}


Recently, the higher order multiplicity moments have gained prominence in high energy physics with a huge hope in pinpointing the QCD critical endpoint (CEP) \cite{endp1} connecting the first order boundary separating the hadronic from the partonic matter at high density with the cross-over boundary at  low density \cite{endp2,Tawfik:2004sw}. It is clear that the hadron resonance gas partition function is an approximation to a non-singular part of the free energy of QCD in the hadronic phase. There are large theoretical uncertainties of its location and even its existence is not fully confirmed, yet \cite{endp3a,endp3b,endp3c}. The characteristic feature of CEP are critical dynamical fluctuations \cite{Tawfik:2010uh,Tawfik:2012zz,Tawfik:2010pt,Tawfik:2005gk,tawDF}. The higher order  moments are conjectured to reflect the large fluctuations associating the hadron-quark phase transition. This was the motivation of a remarkable number of experimental and theoretical studies \cite{endp1,star1,star2,goore}. Recently, various calculations have shown that the higher order moments of the multiplicity distributions of some conserved quantities, such as net-baryon, net-charge, and net-strangeness, are sensitive to the correlation length $\xi$ \cite{endp4,endp4a,corrlenght}, which in turn is related to the higher order moments, themselves. In realistic heavy-ion collisions, the correlation length is found to remain finite.  

Apparently, it should not be an exception that the strongly interacting QCD matter undergoes phase transition(s) at extreme conditions, as almost all matter types suffer from such a critical change as the extreme conditions change. The Lattice QCD calculations predict that a cross-over takes place between the hadronic phase and the Quark Gluon Plasma (QGP) phase, when the temperature exceeds  critical value of $T_c\simeq 150-190~$MeV. As given in Ref. \cite{Tawfik:2004sw,refa1}, depending on the different parameters (for instance, the quark favours and their masses) and on the lattice configurations, lattice QCD assigns different values to $T_c$. Furthermore, the value of $T_c$ depends on the baryon chemical potential $\mu$. With vanishing $\mu$ the lattice QCD calculations \cite{endp5} show that the higher order susceptibilities can be related to the higher order moments of the corresponding multiplicity distributions. They also show non-monotonic behavior near $T_c$. Apparently, the QCD phase diagram including CEP can be mapped out using this characteristic behaviour. In literature, one finds that the QCD phase diagram has been determined using various criteria, such as critical energy density \cite{Karsch:2003vd,Karsch:2003zq,Redlich:2004gp} or line of constant physics \cite{Tawfik:2004sw}. Also the chemical freeze-out curve can be characterized assuming constant physics. For a recent and extensive review, we refer to \cite{jean2006} and the references therein.  

At large chemical potential $\mu$, various calculations based on QCD-based models indicate that the transition from the hadronic phase to the QGP phase is of first order. The endpoint connecting this line with the one of the cross over (at small $\mu$) is likely of second order \cite{endp3c,refa2,endp3a,endp3b,endp3c}. Therefore, the phase transition should be accompanied with large fluctuations in different physical quantities. Experimentally, we study the QCD phase diagram by varying the colliding energy in heavy-ion collisions and determining the critical temperature $T_c$ \cite{endp2}. The possibility of finding QCD CEP has motivated the RHIC beam energy scan program \cite{refa5}. By tuning the collision energy from a center-of-mass energy of $200\;$GeV down to $5\;$GeV, we are able to vary the baryon chemical potential from $\sim 2$ to $\sim500\,$MeV. Nevertheless, there are large theoretical uncertainties of its location and even its existence is not fully confirmed \cite{endp3a,endp3b,endp3c,cepB}. The higher order moments of various physical quantities are conjectured to reflect the non-monotonic behaviour near the QCD CEP. In the present work, we give a systematic study for the higher order moments of the multiplicity distribution and show their ability to represent the fluctuations along the QCD phase diagram. 

Many reasons speak for utilizing the physical resonance gas model (HRG) in predicting the hadron abundances and their thermodynamics. The HRG model seems to provide a good description for the thermal  evolution of the thermodynamic quantities in the hadronic matter~\cite{Karsch:2003vd,Karsch:2003zq,Redlich:2004gp,Tawfik:2004sw,Tawfik:2004vv,Tawfik:2006yq,Tawfik:2010uh,Tawfik:2010pt,Tawfik:2012zz} and has been successfully utilized to characterize the conditions deriving the chemical freeze-out at finite densities~\cite{Tawfik:2005qn,Tawfik:2004ss}. In light of this, HRG can be used in calculating the higher order  moments of particle multiplicity using a grand canonical partition function of an ideal gas with all experimentally observed states up to a certain large mass as constituents. The HRG grand canonical ensemble includes two important features \cite{Tawfik:2004sw}; the kinetic energies and the summation over all degrees of freedom and energies of resonances. On other hand, it is known that the formation of resonances can only be achieved through strong interactions~\cite{Hagedorn:1965st}; {\it Resonances (fireballs) are composed of further resonances (fireballs), which in turn consist of  resonances (fireballs) and so on}. In other words, the contributions of the hadron resonances to the partition function are the same as that of free particles with some effective mass. At temperatures comparable to the resonance half-width, the effective mass approaches the physical one \cite{Tawfik:2004sw}. Thus, at high temperatures, the strong interactions are conjectured to be taken into consideration through including heavy resonances. It is found that hadron resonances with masses up to $2\;$GeV are representing suitable constituents for the partition function ~\cite{Karsch:2003vd,Karsch:2003zq,Redlich:2004gp,Tawfik:2004sw,Tawfik:2004vv,Tawfik:2006yq,Tawfik:2010uh,Tawfik:2010pt,Tawfik:2012zz}. Such a way, the singularity expected at the Hagedorn temperature~\cite{Karsch:2003zq,Karsch:2003vd} can be avoided and the strong interactions are assumed to be considered. Nevertheless, the validity of HRG is limited to temperatures below the critical one, $T_c$.

In this paper, we study the first six non-normalized higher order moments of the particle multiplicity in the hadron resonance gas (HRG) model, which is introduce in section \ref{sec:model}. In section \ref{nonnorlTc}, we suggest an answer to the question: What is needed when going from ''trivial'' lower order to ''sophisticated'' higher order moments? Based on results, general expressions for arbitrary higher moments are deduced, so that it is possible to conclude that going from lower to higher order moments is achievable through a series of lower order moments and correlation functions. The  thermal evolution of the first four normalized moments and their products (ratios) are studied at different chemical potentials $\mu$ in section \ref{sec:norm}. Therefore, it is possible to evaluate them at the chemical freeze-out curve, which is characterized by constant $s/T^3$ at all values of $\mu$, where $s$ and $T$ are entropy density and temperature, respectively. Sections \ref{sec:chemFO} and \ref{sec:cep} are devoted to introduce novel conditions describing the chemical freeze-out curve and define the location of the QCD critical endpoint.   
 

\section{Higher Order Moments of the Particle Multiplicity}
\label{sec:model}

In a grand canonical ensemble, it is straightforward to derive an expression for the pressure. 
In section \ref{nonnorlTc}, we give a list of different moments of particle multiplicity. 
As we move from lower order to higher order multiplicities, certain distribution functions are being added / subtracted. These distribution functions represent higher order correlations. In section \ref{sec:norm}, we list out various normalized higher order moments, the normalization being based on the variance $\sigma$. The products (ratios) of normalized moments are elaborated in section \ref{sec:mult}. 
  
The hadron resonances treated as a free gas~\cite{Karsch:2003vd,Karsch:2003zq,Redlich:2004gp,Tawfik:2004sw,Tawfik:2004vv} are
conjectured to add to the thermodynamic pressure in the hadronic phase (below $T_c$). This statement is valid for free as well as for strongly interacting resonances. 
It has been shown that the thermodynamics of strongly interacting  system can also be approximated to an ideal gas composed of hadron resonances with masses $\le 2~$GeV ~\cite{Tawfik:2004sw,Vunog}. Therefore, the confined phase of QCD, the hadronic phase, is modelled as a non-interacting gas of resonances. The grand canonical partition function reads
\bea
Z(T, \mu, V) &=&\Tr\left[ \exp^{\frac{\mu\, N-H}{T}}\right]
\eea
where $H$ is the Hamiltonian of the system and $T$ is the temperature. The Hamiltonian is given by the sum of the kinetic energies of relativistic Fermi and Bose particles. The main motivation of using this Hamiltonian is that it contains all relevant degrees of freedom of confined and  strongly interacting matter. It includes implicitly the interactions that result in resonance formation. In addition, it has been shown that this model can submit  a quite satisfactory description of particle production in heavy-ion collisions. With the above assumptions the dynamics the partition function can be calculated exactly and be expressed as a sum over 
{\it single-particle partition} functions $Z_i^1$ of all hadrons and their resonances.
\bea
\ln Z(T, \mu_i ,V)&=&\sum_i \ln Z^1_i(T,V)=\sum_i\pm \frac{V g_i}{2\pi^2}\int_0^{\infty} k^2 dk \ln\left\{1\pm \exp[(\mu_i -\varepsilon_i)/T]\right\}
\eea
where $\epsilon_i(k)=(k^2+ m_i^2)^{1/2}$ is the $i-$th particle dispersion relation, $g_i$ is
spin-isospin degeneracy factor and $\pm$ stands for bosons and fermions, respectively.

Before the discovery of QCD, a probable phase transition of a massless pion gas to a new phase of matter was speculated \cite{lsm1}. Based on statistical models like Hagedorn \cite{hgdrn1} and Bootstrap \cite{boots1}, the thermodynamics of such an ideal pion gas is studied, extensively. After the QCD, the new phase of matter is known as quark gluon plasma (QGP). The physical picture was that at $T_c$ the additional degrees of freedom carried by QGP are to be released resulting in an increase in the thermodynamic quantities like energy and pressure densities. The success of hadron resonance gas model in reproducing lattice QCD results at various quark flavours and masses (below $T_c$) changed this physical picture drastically. Instead of releasing additional degrees of freedom at $T>T_c$, the interacting system reduces its effective degrees of freedom at $T<T_c$. In other word, the hadron gas has much more degrees of freedom than QGP.

At finite temperature $T$ and baryon chemical potential $\mu_i $, the pressure of the $i$-th hadron or resonance species reads 
\begin{eqnarray}
\label{eq:lnz1} 
p(T,\mu_i ) &=& \pm \frac{g_i}{2\pi^2}T \int_{0}^{\infty}
           k^2 dk  \ln\left\{1 \pm \exp[(\mu_i -\varepsilon_i)/T]\right\}.
\end{eqnarray}
As no phase transition is conjectured in HRG, summing over all hadron resonances results in the final thermodynamic pressure in the hadronic phase. 

The switching between hadron and quark chemistry is given by the relations between  the {\it hadronic} chemical potentials and the quark constituents; 
$\mu_i =3\, n_b\, \mu_q + n_s\, \mu_S$, where $n_b$($n_s$) being baryon (strange) quantum number. The chemical potential assigned to the light quarks is $\mu_q=(\mu_u+\mu_d)/2$ and the one assigned to strange quark reads $\mu_S=\mu_q-\mu_s$. The strangeness chemical potential $\mu_S$ is
calculated as a function of $T$ and $\mu_i $ under the assumption that the overall
strange quantum number has to remain conserved in heavy-ion collisions~\cite{Tawfik:2004sw}.


\subsection{Non-normalized higher order moments}
\label{nonnorlTc}
As given in the introduction, the higher order moments can be studied in different physical quantities. For example, the higher order moments of charged-particle multiplicity distribution have been predicted four decades ago \cite{gupta72}. Recently, the higher order moments of various  multiplicity distributions have been reported by STAR measurements \cite{star1,star2} and lattice QCD calculations \cite{lqcd1,lqcd1a,lqcd2}. The empirical relevance of the higher order moments to the experimental measurements has been suggested in Ref. \cite{endp1}. Accordingly, the measurement of the correlation length seems to be very much crucial. In a future work, we shall discuss the experimental signatures of the higher order moments. Another ingredient would be the experimental sensitivity for the suggested signatures, which are based on singular behavior of thermodynamic functions. For the time being, we just want to mention that the experimental measurements apparently take place at the final state of the collision, which would mean that the signals have to survive the extreme conditions in such high energy collisions. It has been pointed out that the contribution of the critical fluctuations to the higher order moments is proportional to a positive power of $\xi$. The latter is conjectured to  diverge at QCD CEP and such an assumption is valid in the thermodynamic limit. 

In the present work, we first study the contribution of the hadron resonance mass spectrum in clarifying the interrelations between the various non-normalized and normalized moments. The motivation of using resonance gas is clear \cite{Karsch:2003vd,Karsch:2003zq,Redlich:2004gp}. As introduced about five decades ago, the resonance gas is only valid below $T_c$ \cite{hgdrn1,Karsch:2003vd,Karsch:2003zq,Redlich:2004gp,Tawfik:2004sw,Tawfik:2004vv}. Therefore, the present work is valid in the hadronic phase, only.

For the $i$-th particle, the ''first'' order moment is given by the derivative of $p=-T \partial \ln Z(T,V,\mu_i)/\partial V$ with respect to the dimensionless quantity $\mu_i/T$. When taking into account the antiparticles, we add a negative sign to the chemical potential. The first derivative describes the multiplicity distribution or an expectation operator, which is utilized to estimate the number or multiplicity density 
\bea \label{eq_m1a}
m_1(T,\mu_i ) &=& \pm \frac{g_i}{2\pi^2} T \int_{0}^{\infty}
\frac{e^{\frac{\mu_i - \varepsilon_i}{T}}\; k^2\, dk}{1\pm e^{\frac{\mu_i - \varepsilon_i}{T}}}.
\eea
The ''second'' order moment is known as the variance. It gives the susceptibility of the measurements. 
\bea \label{eq_m2a}
m_2(T,\mu_i ) &=& \pm \frac{g_i}{2\pi^2} T \int_{0}^{\infty}  \frac{e^{\frac{\mu_i -\varepsilon_i}{T}}\; k^2\, dk}{1\pm e^{\frac{\mu_i -\varepsilon_i}{T}}} 
- \frac{g_i}{2\pi^2} T  \int_{0}^{\infty} \frac{e^{2\frac{\mu_i -\varepsilon_i}{T}}\; k^2\, dk }{\left(1\pm e^{\frac{\mu_i -\varepsilon_i}{T}}\right)^2}.
\eea
The ''third'' order moment measures of the lopsidedness of the distribution. As
given in section \ref{sec:norm}, the normalization of third order moment is known as skewness or the asymmetry in the distribution. Skewness tells us about the direction of variation of the data set. 
\bea \label{eq_m3a}
m_3(T,\mu_i ) & = & \pm \frac{g_i}{2\pi^2} T \int_{0}^{\infty}
\frac{e^{\frac{\mu_i -\varepsilon_i}{T}}\; k^2\, dk}{1\pm
  e^{\frac{\mu_i -\varepsilon_i}{T}}} 
- \frac{g_i}{2\pi^2} 3T \int_{0}^{\infty}
\frac{e^{2\frac{\mu_i -\varepsilon_i}{T}}\; k^2\, dk}{\left(1\pm
  e^{\frac{\mu_i -\varepsilon_i}{T}}\right)^2} \nonumber \\
&\pm & \frac{g_i}{2\pi^2} 2T \int_{0}^{\infty}  \frac{e^{3\frac{\mu_i -\varepsilon_i}{T}}\; k^2\, dk}{\left(1\pm e^{\frac{\mu_i -\varepsilon_i}{T}}\right)^3}. 
\eea
In general, the normalization of $r-$th order moment is obtained by dividing it by $\sigma^r$, where $\sigma$ is the standard deviation. The normalization is assumed to remove the brightness dependence.

The ''fourth'' order moment compares the tallness and skinny or shortness and squatness, i.e. shape, of a certain measurement to its normal distribution. It is defined as the uncertainty is an uncertainty or {\it ''the location- and scale-free movement of probability mass from the shoulders of a distribution into its center and tails and to recognize that it can be formalized in many ways''} \cite{kurts1}. 
\bea \label{eq_m4a}
m_4(T,\mu_i ) &=& 
\pm \frac{g_i}{2\pi^2} T \int_0^{\infty} \frac{e^{\frac{\mu_i -\varepsilon_i}{T}}\;
  k^2\, dk}{1\pm e^{\frac{\mu_i -\varepsilon_i}{T}}}
- \frac{g_i}{2\pi^2} 7 T \int_0^{\infty} \frac{e^{2\frac{\mu_i -\varepsilon_i}{T}}\;
  k^2\, dk}{\left(1\pm e^{\frac{\mu_i -\varepsilon_i}{T}}\right)^2} 
\pm \frac{g_i}{2\pi^2} 12 T \int_0^{\infty} \frac{e^{3\frac{\mu_i -\varepsilon_i}{T}}\; k^2\, dk}{\left(1\pm e^{\frac{\mu_i -\varepsilon_i}{T}}\right)^3} \nonumber \\
& & - \frac{g_i}{2\pi^2} 6 T \int_0^{\infty} \frac{e^{4\frac{\mu_i -\varepsilon_i}{T}}\; k^2\, dk}{\left(1\pm e^{\frac{\mu_i -\varepsilon_i}{T}}\right)^4}. 
\eea
The ''fifth'' order moment measures the asymmetry sensitivity of the ''fourth'' order moment.
\bea \label{eq_m5a}
m_5(T,\mu_i ) &=& 
\pm  \frac{g_i}{2 \pi ^2} T \int_0^{\infty }
\frac{e^{\frac{\mu_i -\varepsilon_i}{T}} \; k^2\, dk}{1\pm e^{\frac{\mu_i -\varepsilon_i}{T}}} 
- \frac{g_i}{2 \pi ^2} 15 T \int_0^{\infty } \frac{e^{\frac{2
      (\mu_i -\varepsilon_i)}{T}} \; k^2\, dk}{\left(1\pm e^{\frac{\mu_i -\varepsilon_i}{T}}\right)^2}
\pm \frac{g_i}{2 \pi ^2} 50 T \int_0^{\infty } \frac{e^{\frac{3
      (\mu_i -\varepsilon_i)}{T}} \; k^2\, dk}{\left(1\pm e^{\frac{\mu_i -\varepsilon_i}{T}}\right)^3} \nonumber \\
&  & - \frac{g_i}{2 \pi ^2} 60 T \int_0^{\infty } \frac{e^{\frac{4
      (\mu_i -\varepsilon_i)}{T}} \; k^2\, dk}{\left(1\pm e^{\frac{\mu_i -\varepsilon_i}{T}}\right)^4} 
\pm \frac{g_i}{2 \pi ^2} 24 T \int_0^{\infty } 
\frac{e^{\frac{5 (\mu_i -\varepsilon_i)}{T}}\; k^2\, dk}{\left(1\pm e^{\frac{\mu_i -\varepsilon_i}{T}}\right)^5}. 
\eea
The ''sixth'' order moment is generally associated with compound options.
\bea \label{eq_m6a}
m_6(T,\mu_i ) &=& 
\pm  \frac{g_i}{2 \pi^2} T \int_0^{\infty}
  \frac{e^{\frac{\mu_i -\varepsilon_i}{T}} \; k^2\, dk}{1\pm 
  e^{\frac{\mu_i -\varepsilon_i}{T}}}
- \frac{g_i}{2 \pi^2} 31 T \int_0^{\infty}
  \frac{e^{2\frac{\mu_i -\varepsilon_i}{T}}
  \; k^2\, dk}{\left(1\pm e^{\frac{\mu_i -\varepsilon_i}{T}}\right)^2}
\pm \frac{g_i}{2 \pi^2} 180 T \int_0^{\infty}
  \frac{e^{3\frac{\mu_i -\varepsilon_i}{T}}
  \; k^2\, dk}{\left(1\pm e^{\frac{\mu_i -\varepsilon_i}{T}}\right)^3} \nonumber \\
& & - \frac{g_i}{2 \pi^2} 390 T \int_0^{\infty}
  \frac{e^{4\frac{\mu_i -\varepsilon_i}{T}}
  \; k^2\, dk}{\left(1\pm e^{\frac{\mu_i -\varepsilon_i}{T}}\right)^4} 
\pm \frac{g_i}{2 \pi^2} 360 T \int_0^{\infty}
  \frac{e^{5\frac{\mu_i -\varepsilon_i}{T}}
  \; k^2\, dk}{\left(1\pm e^{\frac{\mu_i -\varepsilon_i}{T}}\right)^5} \nonumber \\
&-& \frac{g_i}{2 \pi^2} 120 T \int_0^{\infty} 
  \frac{e^{6\frac{\mu_i -\varepsilon_i}{T}}
  \; k^2\, dk}{\left(1\pm e^{\frac{\mu_i -\varepsilon_i}{T}}\right)^6}. \hspace*{7mm}
\eea

Thus, from Eqs. (\ref{eq_m1a})-(\ref{eq_m6a}), a general expression for the $r$-th order moment can be deduced 
\bea \label{eq:genralm1}
m_r(T,\mu_i ) &=& \frac{g_i}{2\pi^2}  T \sum_{l=1}^{r} a_{r,l} \int_0^{\infty}
\frac{e^{l \frac{\mu_i -\varepsilon_i}{T}}}{\left(1\pm e^{\frac{\mu_i -\varepsilon_i}{T}}\right)^l} \, k^2\, dk, 
\eea
where the coefficients read
\bea \label{coffs}
a_{r,l} &=& (\pm 1)^l (-1)^{l+1} \left[l\; a_{r-1,l} + (l-1)\; a_{r-1,l-1}\right],
\eea
where $l\leq r$ and $a_{r,l}$ vanishes $\forall\, r<1$. It is obvious that the
coefficients of a certain moment are to be determined from a long chain of all
previous ones. Such a conclusion dates back to about four decades \cite{gupta72}, where it has been shown that the coefficients are related to high order correlation functions. Should this assumption is proven to be valid, then expression (\ref{coffs}) gets a novel interpretation. It seems to sum up the correlation functions up to the $r$-th order. According to \cite{gupta72} and when neglecting the two-particle correlations $C_2$, then the higher order moments read
\bea
m_2 &=& \langle(\delta N)^2\rangle \approx 2 \langle N \rangle, \label{eq:mg2}\\
m_3 &=& \langle(\delta N)^3\rangle \approx 4 \langle N\rangle + C_3, \label{eq:mg3}\\
m_4 &\simeq& \langle(\delta N)^4\rangle = 6m_3 + 3 m_2^2 - 8 m_2 + 8 C_4, \label{eq:mg4}
\eea 
where $\delta N = N-\langle N\rangle$, 
\bea
C_3(p_1,p_2,p_3) &=& \sum_{p_1} \langle p_1p_1p_1\rangle + 3\sum_{p_1<p_2} \langle p_1p_2p_2\rangle +6\sum_{p_1<p_2<p_3} \langle p_1p_2p_3\rangle, \label{eq:r2}\\
C_4(p_1,p_2,p_3,p_4) &=& \sum_{p_1<p_2} \langle p_1p_2p_2p_2\rangle + \sum_{p_1<p_2<p_3} \langle p_1p_2p_3p_3\rangle + 8 \sum_{p_1<p_2<p_3<p_4} \langle p_1p_2p_3p_4\rangle,  \label{eq:r3}
\eea
and $p_i$ is the $i-$th particle. 
To the second order moment $m_2$ we have to add the effects of the two particle correlation function, $2 \sum C_2$. The third order moment $m_3$ gets approximately three times this amount. The three and four particle correlations, Eqs. (\ref{eq:r2}) and (\ref{eq:r3}), appear first in third and fourth moment, Eq. (\ref{eq:mg4}) and (\ref{eq:r2}), respectively. 

When ignoring these high order correlations, we are left with the widely used two particle correlations. In the  present work, we restrict the discussion to this type, only. It is of great interest, as it is accessible experimentally and achievable, numerically. The two particle correlations are suggested as a probe for the bulk QCD medium, energy loss, medium response, jet properties and intensity interferometry \cite{tp1,tp2,tp3,tp4,tp5,tp6,tp7}. In addition to this list of literature, the comprehensive review \cite{physrep} can be recommended.

\begin{figure}[htb]
\includegraphics[angle=-90,width=8.cm]{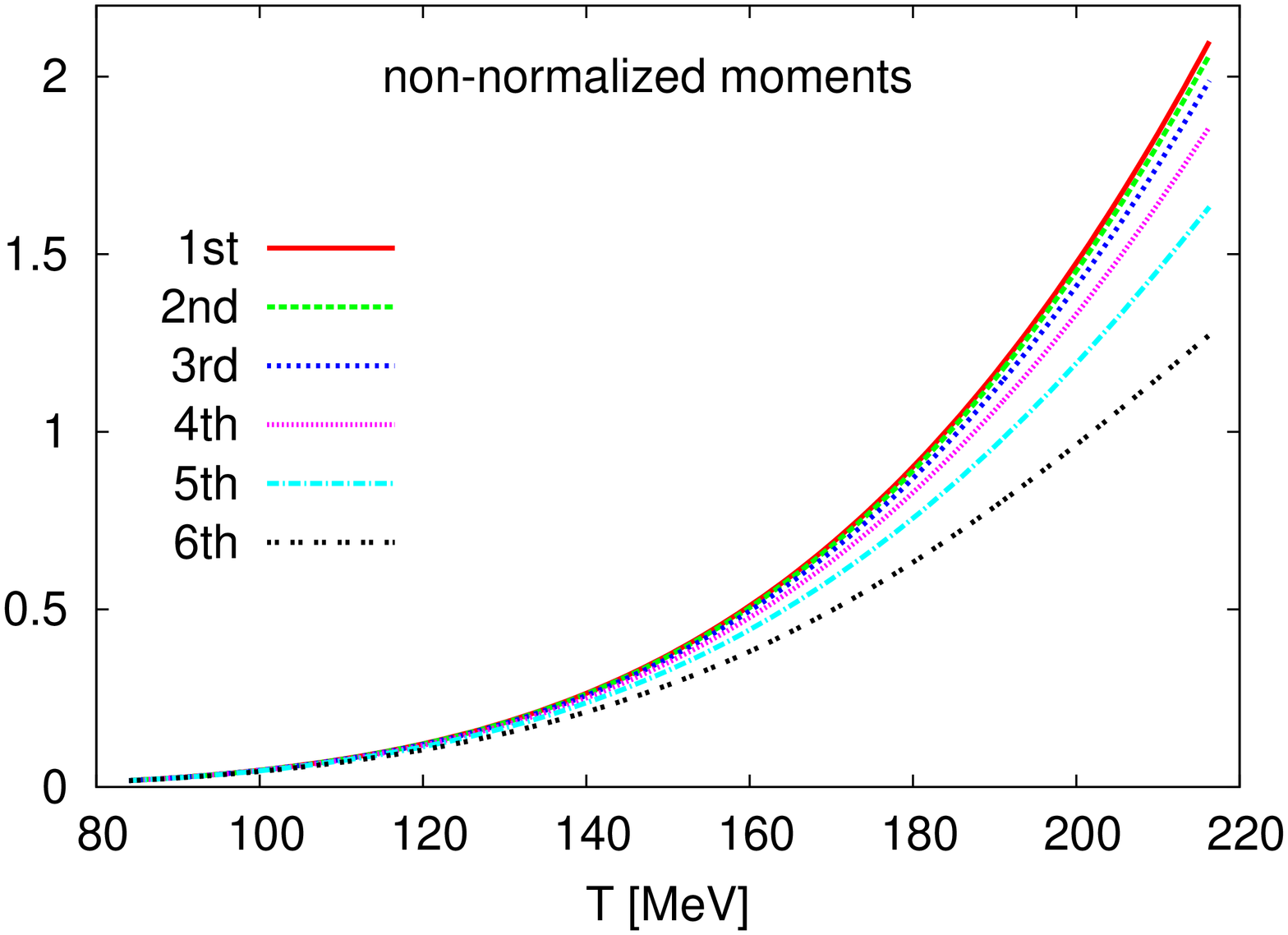}
\includegraphics[angle=-90,width=8cm]{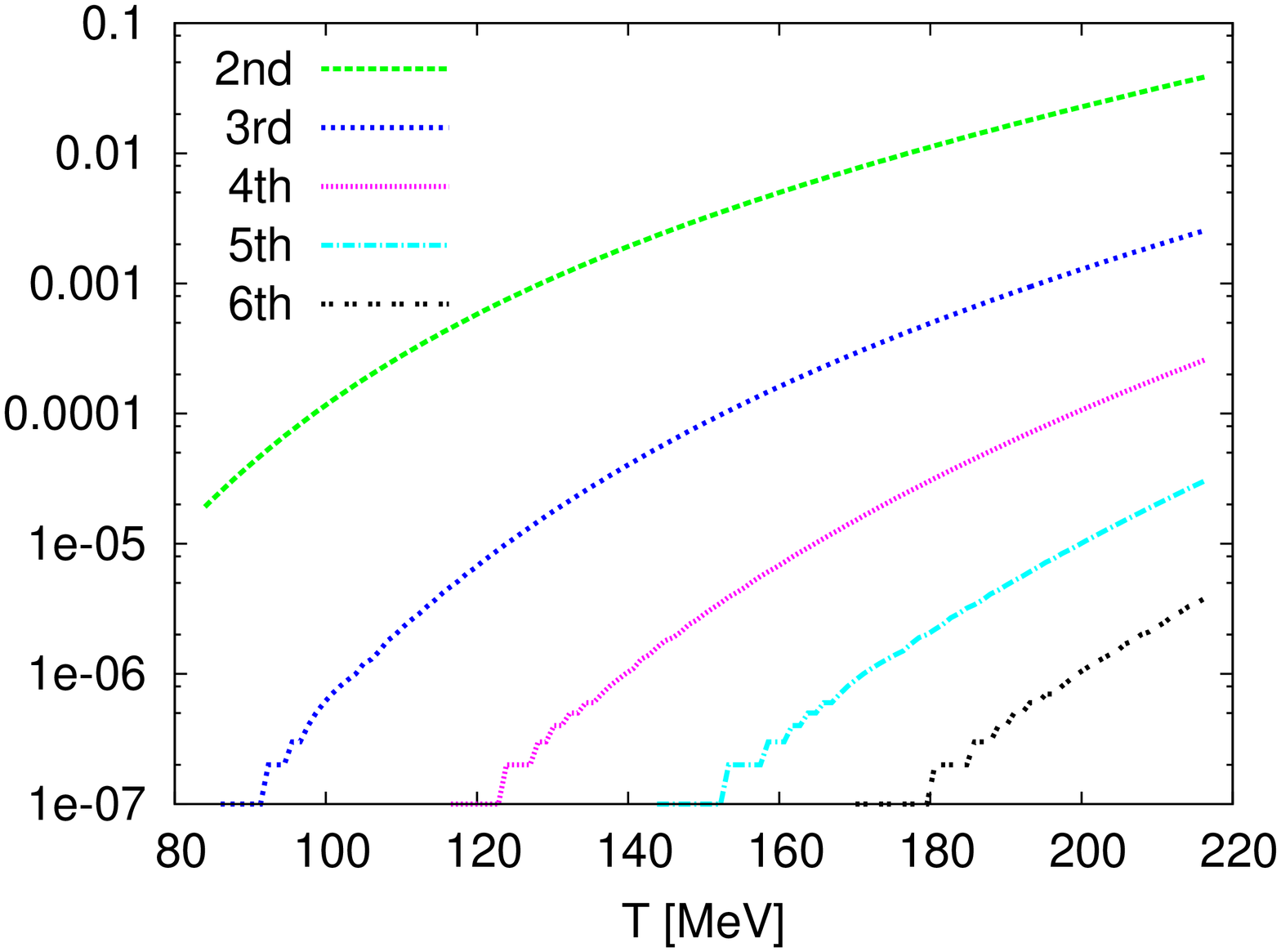}
\caption{Left panel depicts dimensionless non-normalized moments in dependence on the temperature.  Right panel shows the evolution of the so-called correlation parts (the integrals appearing in Eqs. (\ref{eq_12b})-(\ref{eq_m6b})) with the temperature. The right panel shows  the  new coefficients that resulted in when going from lower to higher order moments of multiplicities in a log-scale.}
\label{fig:hrg1} 
\end{figure}

Taking into account the particle multiplicities, then expressions (\ref{eq:mg2}), (\ref{eq:mg3}) and (\ref{eq:mg4}) can be re-written as follows.
\bea
m_2 = \langle(\delta N)^2\rangle &\simeq& \langle N^2\rangle - \langle N\rangle^2, \\
m_3 = \langle(\delta N)^3\rangle &\simeq& \langle N^3\rangle - \langle N^2\rangle \langle N\rangle + 2 \langle N^3\rangle^3, \\
m_4 = \langle(\delta N)^4\rangle - 3 \langle(\delta N)^2\rangle^2 
   &\simeq & \langle\langle N\rangle^4\rangle - 2 \langle\langle N^2\rangle^2\rangle - 5 \langle\langle N^2\rangle\rangle^2 + 6 \langle\langle N\rangle^2\rangle\, \langle\langle N^2\rangle\rangle.
\eea
As mentioned in section \ref{nonnorlTc}, the experimental signals for the higher order moments shall be discussed in a future work.  

As given in Eq. (\ref{eq:genralm1}), the dependence of first six non-normalized moments on lower ones can explicitly be deduced as follows. 
\bea 
m_1 &=& \frac{g_i}{2\pi^2} T \int_{0}^{\infty}  \frac{e^{\frac{\mu_i -\varepsilon_i}{T}}\; k^2\, dk}{1 \pm e^{\frac{\mu_i -\varepsilon_i}{T}}}, \label{eq_12b} \\
m_2 &=&  m_1 - \frac{g_i}{2\pi^2} T  \int_{0}^{\infty}  \frac{e^{2\frac{\mu_i -\varepsilon_i}{T}}\; k^2\, dk}{\left(1 \pm e^{\frac{\mu_i -\varepsilon_i}{T}}\right)^2}, \label{eq_m2b} \\
m_3 &=&  -2 m_1 + 3 m_2 + 2 \frac{g_i}{2\pi^2} T \int_{0}^{\infty}  \frac{ e^{3\frac{\mu_i -\varepsilon_i}{T}}\; k^2\, dk}{\left(1 \pm e^{\frac{\mu_i -\varepsilon_i}{T}}\right)^3}, \label{eq_m3b}\\
m_4 &=& 6 m_1 -11 m_2 + 6 m_3
-6 \frac{g_i}{2\pi^2} T \int_0^{\infty} \frac{e^{4\frac{-\epsilon_i +\mu_i}{T}}\; k^2\, dk}{\left(1 \pm e^{\frac{-\epsilon_i +\mu_i }{T}}\right)^4}, \label{eq_m4b}\\
m_5 &=& -24 m_1 + 50 m_2 - 35 m_3 + 10 m_4
+ 24 \frac{g_i}{2\pi^2} T \int_0^{\infty} \frac{e^{5\frac{-\epsilon_i +\mu_i}{T}}\; k^2\, dk}{\left(1 \pm e^{\frac{-\epsilon_i +\mu_i }{T}}\right)^5}, \label{eq_m5b}\\
m_6 &=& 120 m_1 - 274 m_2 + 225 m_3 - 85 m_4 + 15 m_5 - 120 \frac{g_i}{2\pi^2} T \int_0^{\infty} \frac{e^{6\frac{-\epsilon_i +\mu_i}{T}}\; k^2\, dk}{\left(1 \pm e^{\frac{-\epsilon_i +\mu_i }{T}}\right)^6}. \label{eq_m6b}
\eea
Naively spoken, we conclude that raising lower order moments to higher ones is achievable through a series of all lower order moments and an additional term reflecting the correlations, themselves \cite{gupta72}. From Eqs. (\ref{eq_12b})-(\ref{eq_m6b}), the additional terms are proportional to $\langle N\rangle^r$. Apparently, they are  essential in order to judge whether going from lower to higher order moments would make the signatures of dynamical fluctuations clearer than when excluding them. 

The first and second terms can be generalized. Then, a general expression would read
\bea\label{eq:genmm}
m_r=(-)^{r-1}(r-1)c_{m_1}^{r-1} m_1 - \left[(r-1)c_{m_2}^{r-1} + (r-2)!\right] m_2 +{\cal O}(m_{>2}),
\eea
where $c_{m_{r}}$ is the coefficient of the $r$-th moment. The last term in Eq. (\ref{eq:genmm}) can be elaborated when higher order moments are calculated. The latter are essential in order to find out a clear pattern.

The figure \ref{fig:hrg1} presents the results of the higher order moments as calculated in the HRG model. As introduced previously, raising the orders of multiplicity moments results in new coefficients and new integrals. The earlier are partly characterized in Eq. (\ref{eq:genmm}). 
The dimensionless non-normalized moments are given in dependence on the temperature $T$ in the left panel of Fig. \ref{fig:hrg1}. The new coefficients that resulted in when going from lower to higher order moments are depicted in the right panel. The integrals can be compared with the correlation functions \cite{gupta72}. We find that these integrals are related to $\langle N \rangle^r$, where $r$ is the order of the moment. It is obvious that successive moments have a difference of about one order of magnitude. Therefore, higher order $\langle N\rangle^r$ can be disregarded with reference to their vanishing contributions.

In order to draw a picture about the contributions of the integrals appearing in Eqs. (\ref{eq_12b})-(\ref{eq_m6b}), the phase space integral can be replaced by a series representation. For the $i$-th particle species, the pressure can be expressed as
\bea \label{eq:bssl1}
P(T, \mu_i) &=&  \frac{g_i}{2\, \pi^2}\, T^4\, \sum_{n=1}^{\infty}\,
(\pm)^{n+1}\, e^{n\,\frac{\mu_i}{T}}\, \left(n\, \frac{m_i}{T}\right)^2 \,
n^{-4}\, \text{K}_2\left(n\, \frac{m_i}{T}\right),
\eea
where $\text{K}_2$ is modified Bessel function of the second kind and the factor $(\pm)^{n+1}$ represents fermions and bosons,
respectively. In deriving previous expression, we expand the functions
appearing in Eq. (\ref{eq_m1a}), for instance. The resulting expression is
applicable as long as $\mu_i<m_i$ is valid.
Then, the derivatives with respect to $\mu/T$ leads to
\bea \label{eq:bssl2}
m_r(T,\mu_i) &=& \frac{g_i}{2\, \pi^2}\, T^4\, \sum_{n=1}^{\infty}\,  (\pm)^{n+1}\, e^{n\,
 \frac{\mu_i}{T}}\, \left(n\, \frac{m_i}{T}\right)^2 \,
n^{r-4}\, \text{K}_2\left(n\, \frac{m_i}{T}\right),
\eea 
which is valid for all order moments of order $r\geq 1$.  

For completeness, we mention here that taking into consideration one particle and its antiparticle  is not rare in literature. In this case, the expression (\ref{eq:bssl1}) can be re-written as
\bea \label{eq:PTmu}
P(T, \mu_i) &=&  \frac{g_i}{\pi^2}\, T^4\, \sum_{n=1}^{\infty}\,
(\pm)^{n+1}\, \left(n\, \frac{m_i}{T}\right)^2 \,
n^{-4}\, \text{K}_2\left(n\, \frac{m_i}{T}\right)\, \text{cosh}\left(n\,\frac{\mu_i}{T}\right),
\eea
and accordingly,
\bea \label{eq:bssl2b}
m_r(T,\mu_i) &=& \frac{g_i}{2\, \pi^2}\, T^4\, \sum_{n=1}^{\infty}\,  (\pm)^{n+1}\, \left(n\, \frac{m_i}{T}\right)^2 \, n^{-4}\, \text{K}_2\left(n\, \frac{m_i}{T}\right)\, \frac{\partial^r}{\partial \mu^r} \text{cosh}\left(n\, \frac{\mu_i}{T}\right).
\eea 
Comparing Eq. (\ref{eq:bssl2}) with Eq. (\ref{eq:bssl1}), we conclude that calculating
higher order moments is associated with a gradual change in $n$. Expression
(\ref{eq:genralm1}) reflects almost the same behaviour, where higher order moments
are associated with change in the coefficients.


\subsection{Normalized higher order moments}
\label{sec:norm}

\begin{figure}[htb]
\includegraphics[angle=-90,width=5.5cm]{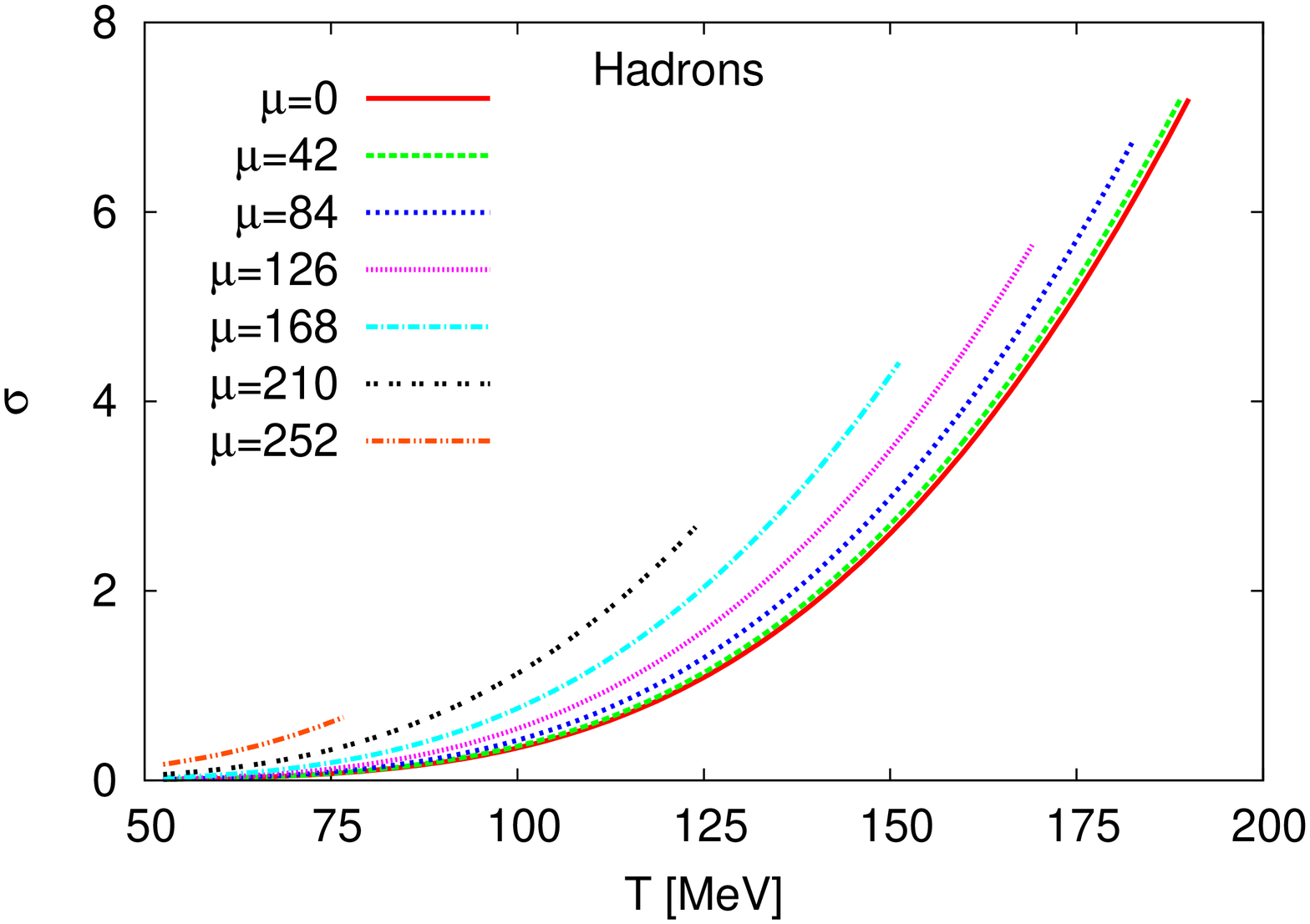}
\includegraphics[angle=-90,width=5.5cm]{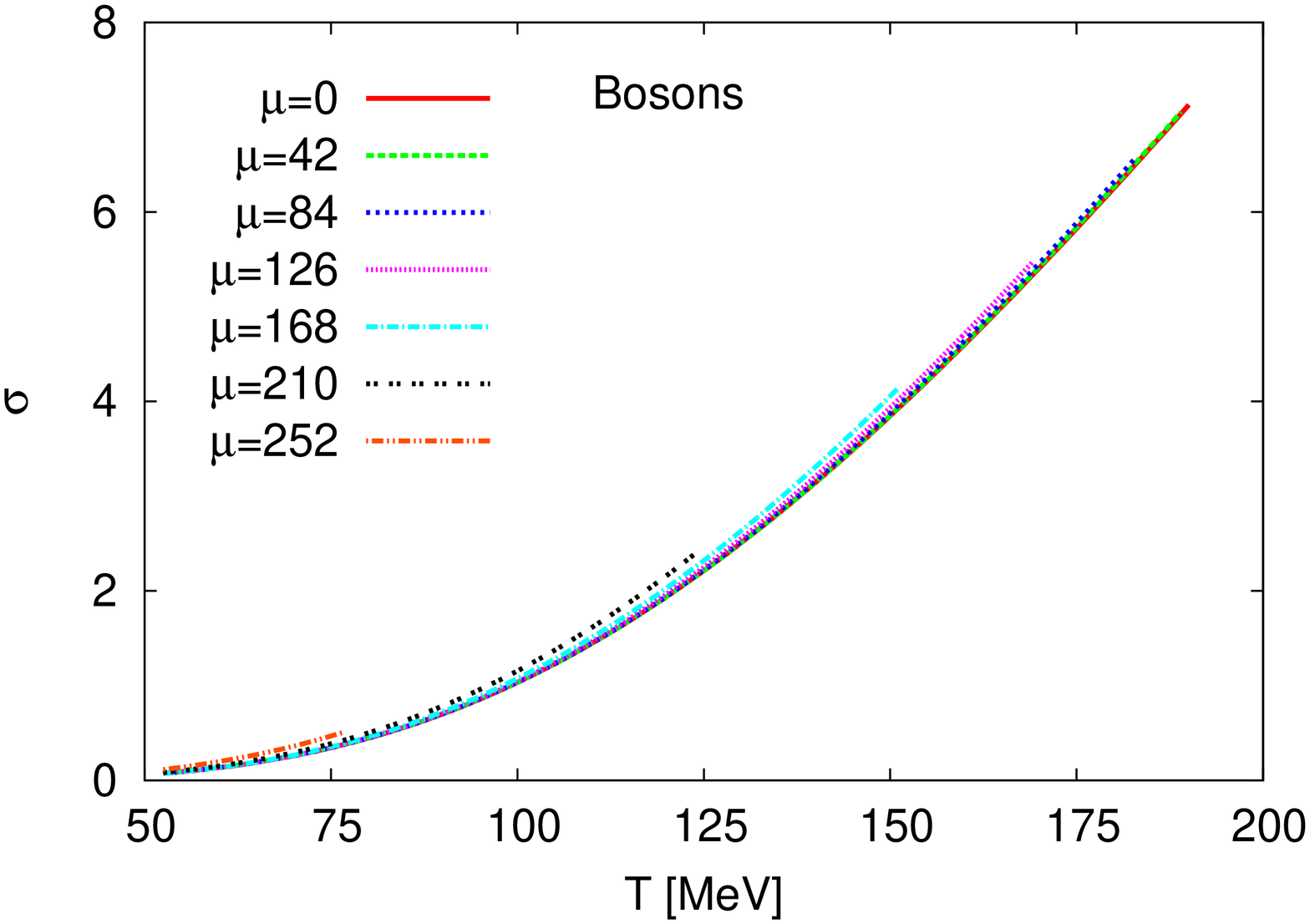}
\includegraphics[angle=-90,width=5.5cm]{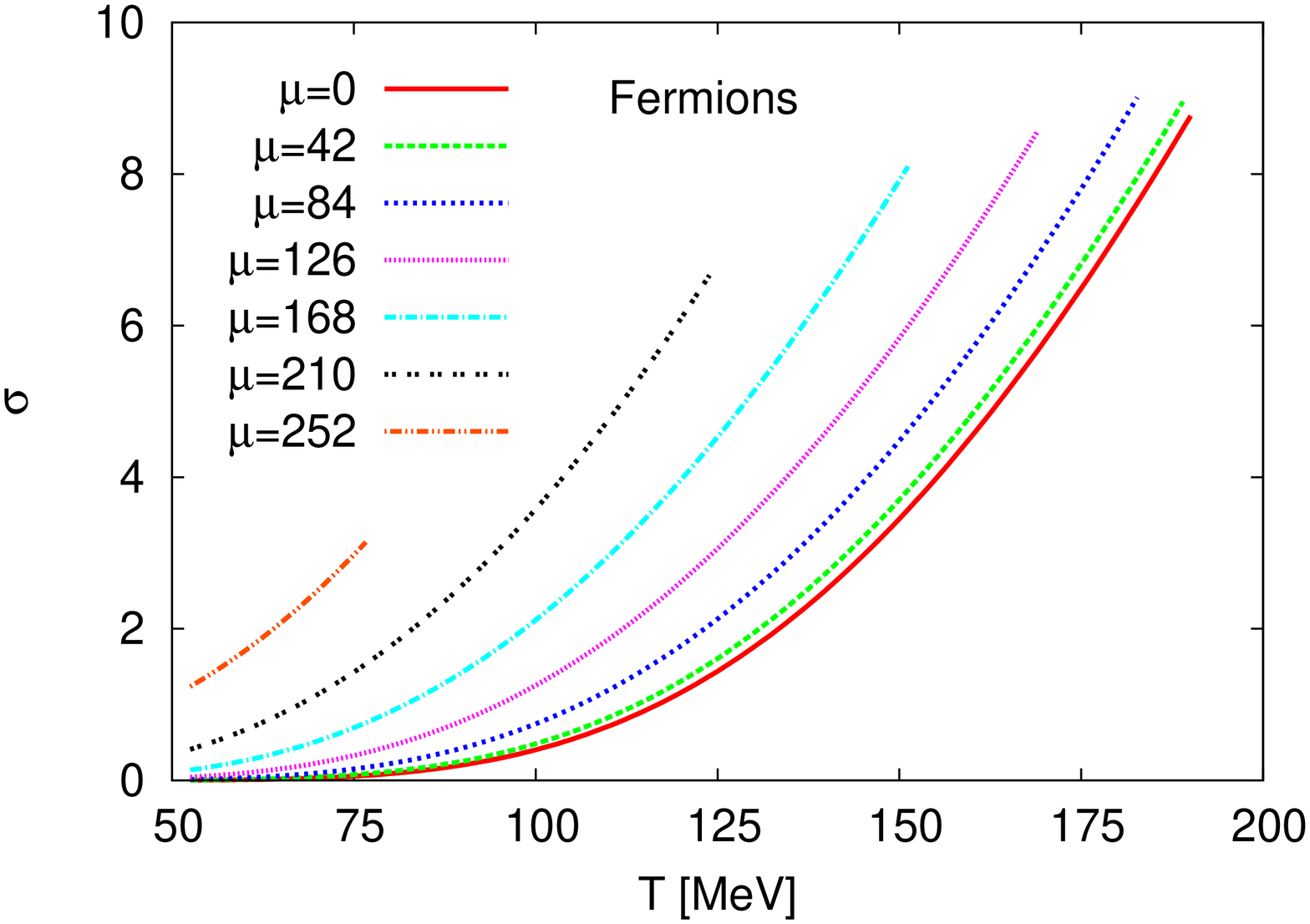}
\caption{Results for  $\sigma$ in hadronic (left), bosonic (middle) and fermionic (right) resonances are given in dependence on $T$ for various baryon chemical potentials (given in MeV).}
\label{fig:sigmaaa} 
\end{figure}

The normalization of higher order moments which can be deduced through derivatives with respect to the chemical potential $\mu$ of given charges, apparently gives additional insights about the properties of higher order moments. From statistical 
point of view, the normalization is done with respect to the standard deviation $\sigma$, which is be related to $\xi$. Therefore, it provides with a tool to relate moments with various orders to the experimental measurement. The susceptibility of the distribution give a measure for $\sigma$. 
Should we are interested on the multiplicity, then the susceptibility is simply 
given by the derivative of first order moments with respect to $\mu$. It has been shown that the susceptibility is related to $\sim\xi^2$ \cite{endp1}. The results of  $\sigma$ in hadronic, bosonic and fermionic resonances are calculated at different $\mu$ and given in Fig. \ref{fig:sigmaaa}. As discussed earlier, the strange quantum number has to remain conserved in high energy collisions. As per the standard model, this is one of the global symmetries in strong interactions. The procedure of keeping strange degrees of freedon conserved in HRG is introduced in Ref. \cite{Tawfik:2004sw}. This is the origin of the $\mu$-dependence, especially in the bosonic gas. Although, baryon chemical potential $\mu$ vanishes per definition, the chemical potential associated with strange quark $\mu_S$ remains finite. The dependence of $\mu_S$ on $\mu$ is depicted in Fig. \ref{fig:mub_mus}. Another feature in these calculations is the assumption that the freeze-out boundary is determined by constant $s/T^3$, where $s$ is the entropy density \cite{Tawfik:2005qn}. At chemical freeze-out boundary, the dependence of $\sigma$ on $\mu$ is given in Fig. \ref{fig:S2n2}.

\begin{figure}[htb]
\includegraphics[angle=-90,width=5.5cm]{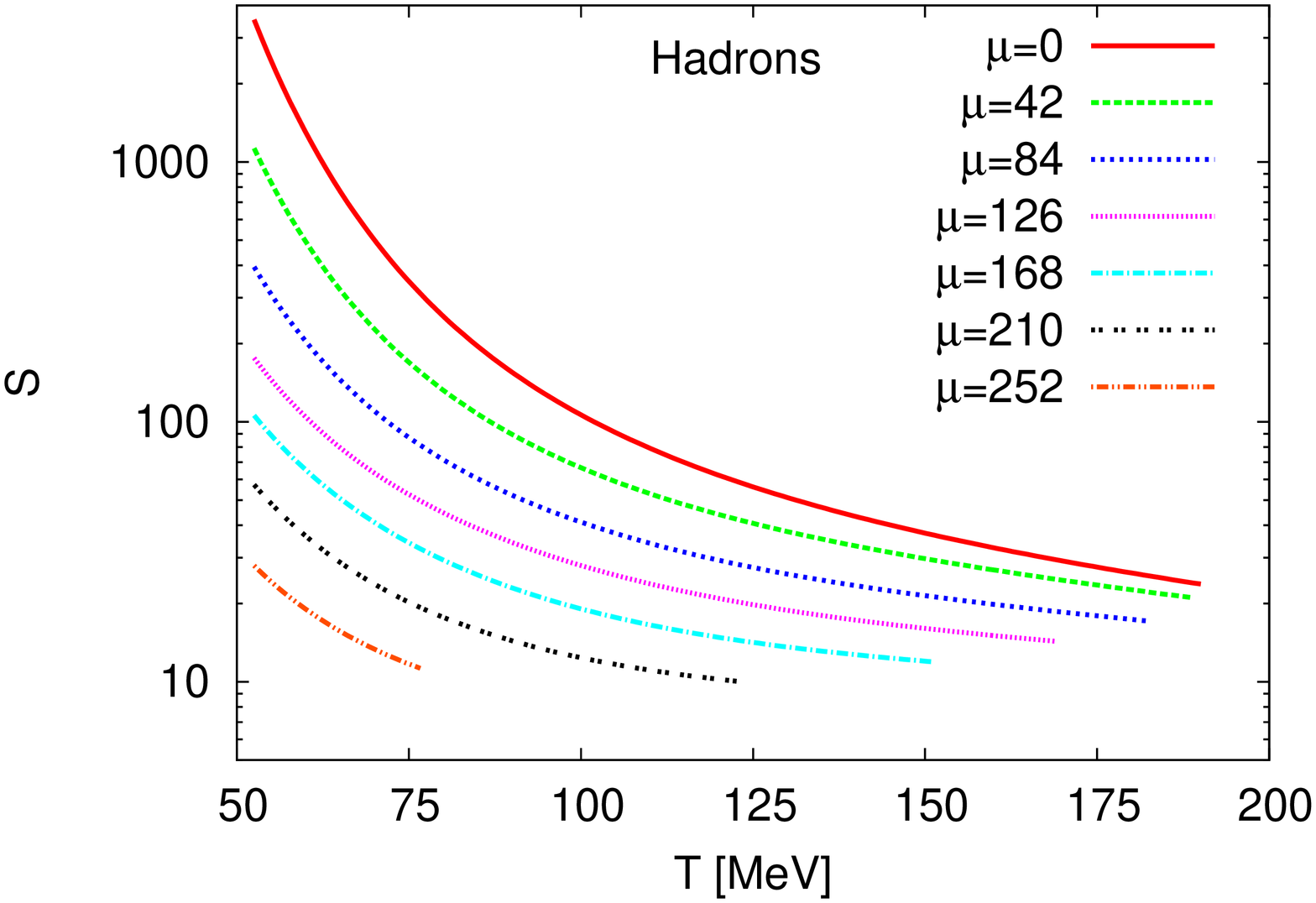}
\includegraphics[angle=-90,width=5.5cm]{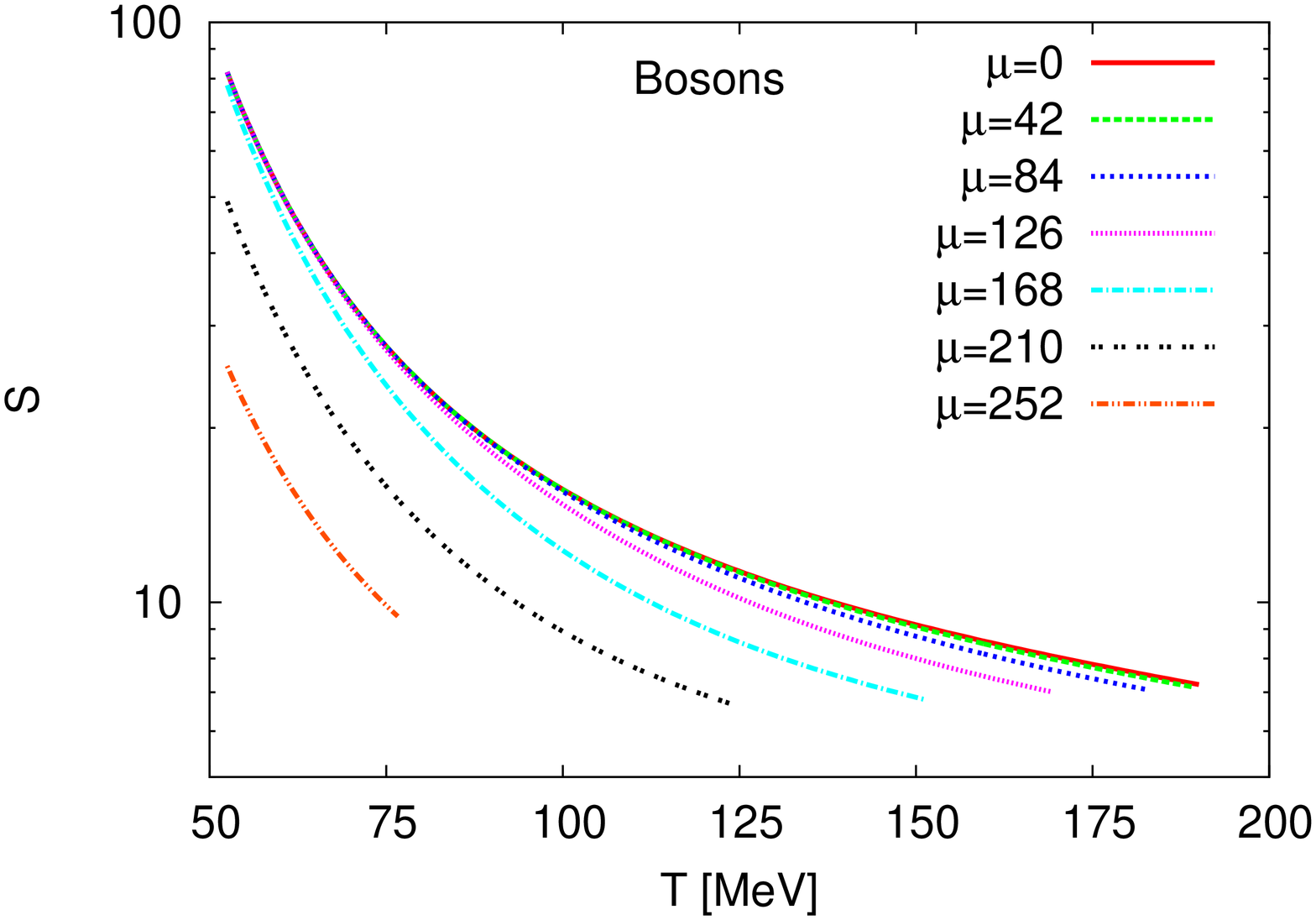}
\includegraphics[angle=-90,width=5.5cm]{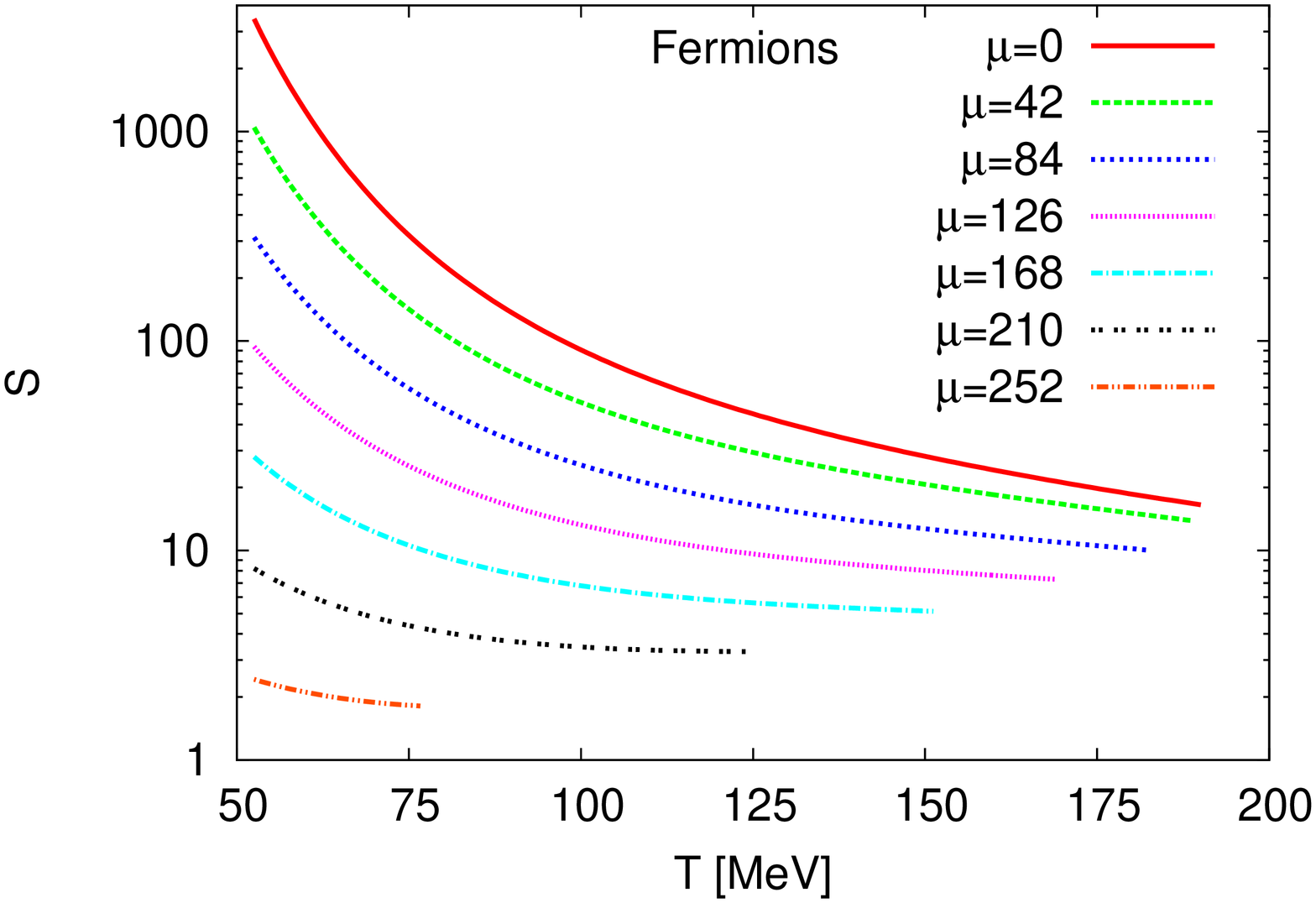}
\caption{At various baryon chemical potentials (given in MeV), the skewness $S$  for hadronic (left), bosonic (middle) and fermionic (right) resonance gas is given as function of $T$.}
\label{fig:S} 
\end{figure}

The ''third'' order moment normalized to $\sigma^3$ is known as 'skewness''. For standard Gaussian distribution, the skewness is obviously vanishing. Therefore, the skewness is an ideal quantity probing the non-Gaussian fluctuation feature as expected near $T_c$. The QCD CEP is conjectured to be sensitive to skewness. Experimentally, it has been shown that 
the skewness $S$ is related to $\sim\xi^{4.5}$ \cite{endp4}. The skewness for bosonic and fermionic resonance gas, respectively, reads
\bea 
S_b &=& -\frac{1}{2}\,\frac{\pi}{\sqrt{g_i}}\, T^{3/2}\, \frac{\int_0^{\infty} \,  
\text{csch}\left[\frac{\varepsilon_i - \mu_i }{2\, T}\right]^4\, \text{sinh} \left[\frac{\varepsilon_i - \mu_i }{T}\right] \; k^2 \, dk}
{\left[\int_0^{\infty} \left(\text{cosh} \left[\frac{\varepsilon_i-\mu_i }{T}\right]-1\right)^{-1}\; k^2 \, dk \right]^{3/2}}, \label{eq:Ssb}\\
S_f &=& 8\,\frac{\pi}{\sqrt{g_i}}\, T^{3/2}\, \frac{\int_0^{\infty} \,  
\text{csch}\left[\frac{\varepsilon_i - \mu_i }{T}\right]^3\, \text{sinh}
\left[\frac{\varepsilon_i - \mu_i }{2\, T}\right]^4 \; k^2 \, dk}
{\left[\int_0^{\infty} \left(\text{cosh} \left[\frac{\varepsilon_i-\mu_i }{T}\right]+1\right)^{-1}\; k^2 \, dk \right]^{3/2} }. \label{eq:Ssf}
\eea
At different chemical potentials, the skewness $S$ is calculated in dependence on $T$ and given in Fig. \ref{fig:S}. It is worthwhile to notice that the values of $S$ in the fermionic resonance gas is much larger than the $S$-values in the bosonic resonance gas, especially at small chemical potentials. But at large chemical potential, this observation is exactly opposite. The value of $\mu$, at which the large (or small) skewness is flipped, seems to reflect the nature of phase transition, where the fermionic and/or bosonic degrees of freedom turns out to be undistinguishable. Such an observation shall be utilized in positioning the QCD CEP, section \ref{sec:chemFO}.

The normalization of $4-$th order moment is known as heteroskedacity or kurtosis. It means varying volatility or more accurately, varying variance.  Actually, the kurtosis is given by normalized $4-$th order moment minus $3$. The subtraction of $3$, which arises from the Gaussian distribution,  is usually omitted \cite{kurts1,kurts2,kurts3}.  Therefore, the kurtosis is an ideal quantity for probing the non-Gaussian fluctuation feature as expected near $T_c$ and CEP. A sign change of skewness or kurtosis is conjectured to indicate that the system crosses the phase boundary \cite{endp5,endp6}. As HRG is valid below $T_c$, the sign change is not accessible. It has been shown that kurtosis $\kappa$ is related to $\sim\xi^{7}$ \cite{endp4}.
\bea
\kappa_b &=& -\frac{\pi^2}{g_i}\, T^3\, \frac{\int_0^{\infty}
  \left\{\text{cosh}\left[\frac{\varepsilon_i - \mu_i }{T}\right] + 2\right\} 
\text{csch}\left[\frac{\varepsilon_i - \mu_i }{2\, T}\right]^4 \; k^2 \, dk}
{\left[\int_0^{\infty} \left(1-\text{Cosh}\left[\frac{\varepsilon_i-\mu_i }{T}\right]\right)^{-1}\; k^2 \, dk\right]^{2} } - 3,  \label{eq:Kkb}\\
\kappa_f &=& \frac{\pi^2}{g_i}\, T^3\, \frac{\int_0^{\infty}
  \left\{\text{cosh}\left[\frac{\varepsilon_i - \mu_i }{T}\right] - 2\right\} 
\text{sech}\left[\frac{\varepsilon_i - \mu_i }{2\, T}\right]^4 \; k^2 \, dk}
{\left[\int_0^{\infty} \left(\text{cosh}\left[\frac{\varepsilon_i-\mu_i }{T}\right]+1\right)^{-1}\; k^2 \, dk\right]^{2} } - 3.  \label{eq:Kkf}
\eea


\subsection{Products of higher order moments}
\label{sec:mult}

There are several techniques to scale the correlation functions. The survey system's optional statistics module represents the most common technique i.e., Pearson or product moment correlation. This module includes the so-called partial correlation which seems to be useful when the relationship between two variables is to be highlighted, while effect of one or two other variables can be removed.
In the present work, we study the  products of higher order moments of the distributions of conserved quantities. The justification of this step is that certain products can be directly connected to the corresponding susceptibilities in Lattice QCD simulation and related to long range correlations \cite{endp5,qcdlike,latqcd1}. Seeking for simplicity, we start with the Boltzmann approximation.

\label{sec:besslB}

\subsubsection{Modified Bessel function: Boltzmann statistics}
\label{sec:bessel}

When relativistic momentum integrals are replaced by summation over modified
Bessel functions, the products of higher order moments in Boltzmann approximation read
\bea
\frac{\sigma^2}{\langle N\rangle} &=& \frac{\sum_{n=1}^{\infty}  (\pm)^{n+1}\, e^{n\frac{\mu_i}{T}}
  \left(n\frac{m_i}{T}\right)^2 n^{-2} K_2\left(n\frac{m_i}{T}\right)}{\sum_{n=1}^{\infty}  (\pm)^{n+1} e^{n\frac{\mu_i}{T}}
  \left(n\frac{m_i}{T}\right)^2 n^{-3} K_2\left(n\frac{m_i}{T}\right)}, \label{eq:k21}\\
S\, \sigma &=& \frac{\sum_{n=1}^{\infty} (\pm)^{n+1} e^{n\frac{\mu_i}{T}}
  \left(n\frac{m_i}{T}\right)^2 n^{-1}  K_2\left(n\frac{m_i}{T}\right)}{\sum_{n=1}^{\infty} \, (\pm)^{n+1} e^{n\frac{\mu_i}{T}}
  \left(n\frac{m_i}{T}\right)^2 n^{-2} K_2\left(n\frac{m_i}{T}\right)}, \label{eq:k22}\\
\kappa\, \sigma^2 &=& \frac{\sum_{n=1}^{\infty} (\pm)^{n+1} e^{n\frac{\mu_i}{T}}
  \left(n\frac{m_i}{T}\right)^2 K_2\left(n\frac{m_i}{T}\right)}{\sum_{n=1}^{\infty} (\pm)^{n+1} e^{n\frac{\mu_i}{T}}
  \left(n\frac{m_i}{T}\right)^2 n^{-2} K_2\left(n\frac{m_i}{T}\right)} - 3
\frac{g_i}{2 \pi^2} T^4 \sum_{n=1}^{\infty} (\pm)^{n+1}\,
e^{n\frac{\mu_i}{T}}\, \left(n\frac{m_i}{T}\right)^2 n^{-2}\, K_2\left(n\frac{m_i}{T}\right), \label{eq:k23} \\
\frac{\kappa\, \sigma}{S} &=& \frac{2 \pi^2 \sum_{n=1}^{\infty} 
  (\pm)^{n+1} e^{n\frac{\mu_i}{T}} \left(n\frac{m_i}{T}\right)^2 
  K_2\left(n\frac{m_i}{T}\right) - 3 g_i T^4 \left(\sum_{n=1}^{\infty}
  (\pm)^{n+1} 
e^{n\frac{\mu_i}{T}} \left(n\frac{m_i}{T}\right)^2 n^{-2} K_2\left(n\frac{m_i}{T}\right)\right)^2}{2\,\pi^2\sum_{n=1}^{\infty} (\pm)^{n+1}
  e^{n\frac{\mu_i}{T}} \left(n\frac{m_i}{T}\right)^2 n^{-1} K_2\left(n\frac{m_i}{T}\right)}. \hspace*{10mm} \label{eq:k24}
\eea
The origin of the second term in all expressions containing $\kappa$ is obvious. Expressions (\ref{eq:k21})-(\ref{eq:k24}) justify the conclusions in \cite{KF2010} that in Boltzmann approximation 
\bea
\frac{\sigma^2}{\langle N\rangle} &=& S\, \sigma \simeq 1,
\eea
while
\bea
\kappa\, \sigma^2 = \frac{\kappa\, \sigma}{S} &\simeq& 1 - 3 \frac{g_i}{2 \pi^2} T^4 \text{exp}\left[\frac{\mu_i}{T}\right]\, \left(\frac{m_i}{T}\right)^2 \, K_2\left(\frac{m_i}{T}\right).
\eea
These expressions are valid in the final state, which can be characterized by chemical and thermal freeze-out. In other words, these expressions are functions of the chemical potential. 
Using relativistic momentum integrals (see section \ref{sec:rmi}) shows that the products of moments result in a constant dependence on $T$ \cite{KF2010}.


\subsubsection{Relativistic Momentum Integrals; quantum statistics}
\label{sec:rmi}

The fluctuations of conserved quantities are assumed to be sensitive to the structure of the hadronic system in its final state. As mentioned above, crossing the phase boundary or passing through critical endpoint is associated with large fluctuations. Most proposed fluctuation of observables are variations of second order moments of the distribution, such as particle ratio \cite{tawDF} and charged dynamical measurement \cite{chargeD}. Then, the fluctuations are approximately related to $\xi^2$ \cite{corrlenght}.

\begin{figure}[htb]
\includegraphics[angle=-90,width=5.5cm]{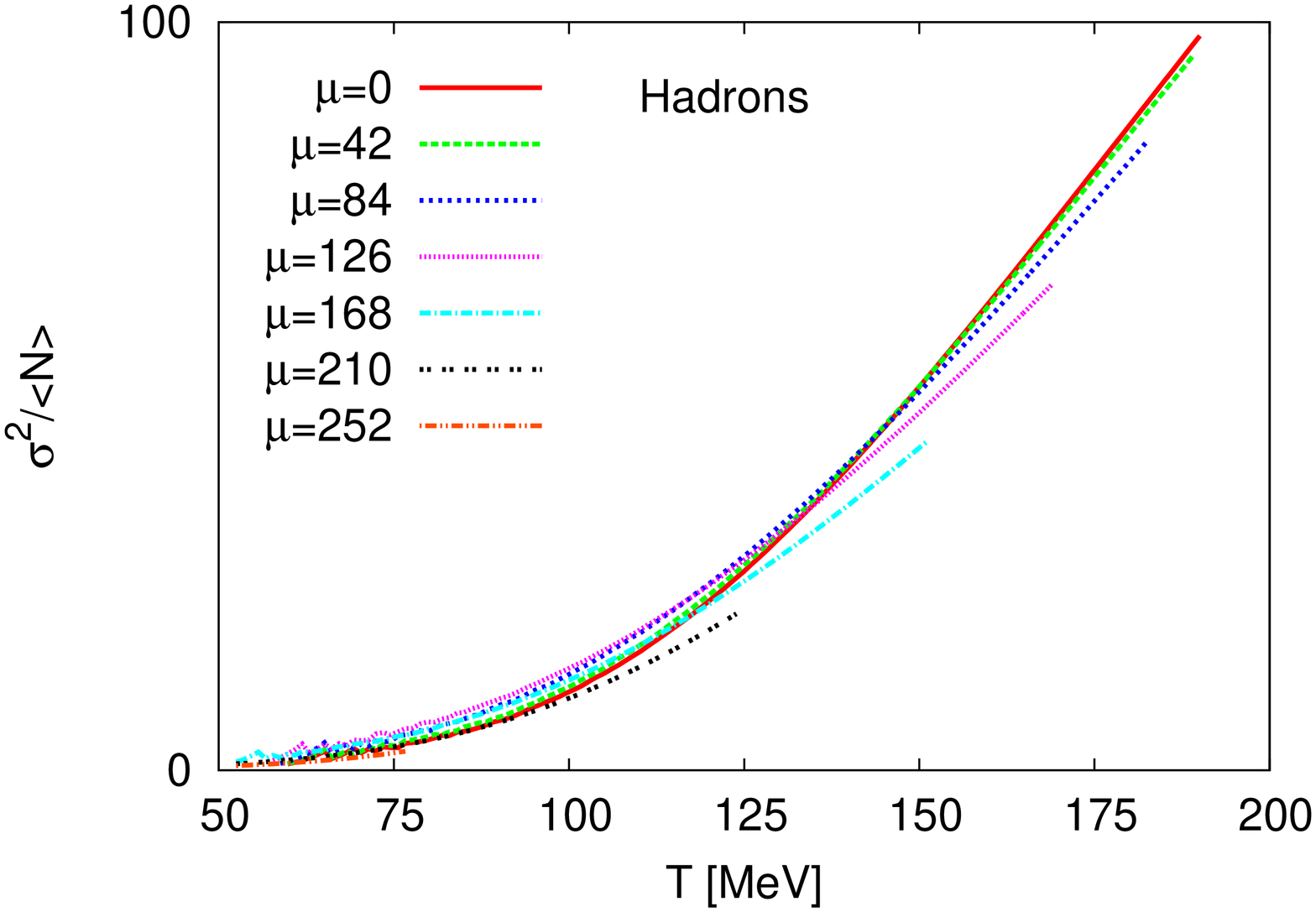}
\includegraphics[angle=-90,width=5.5cm]{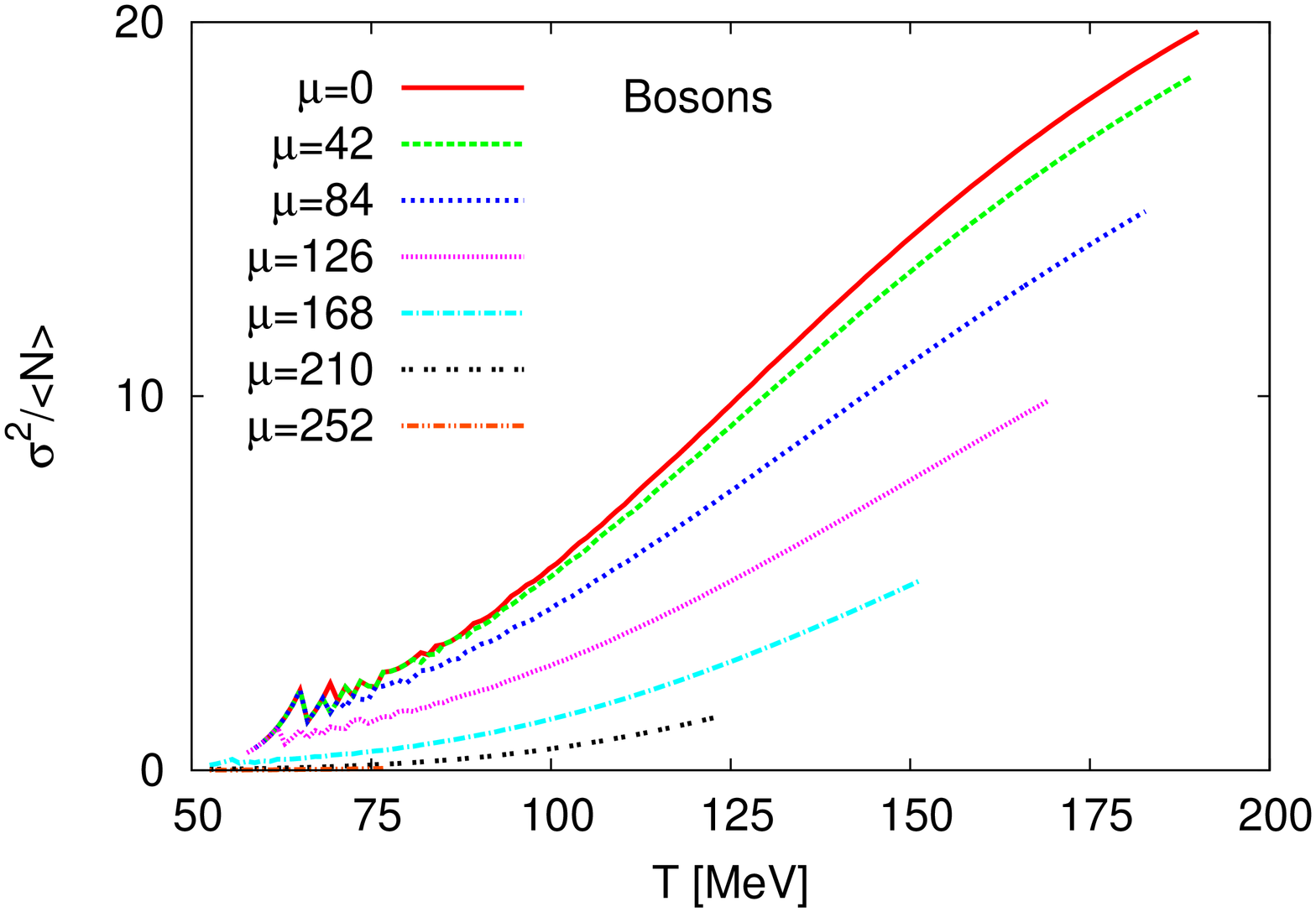}
\includegraphics[angle=-90,width=5.5cm]{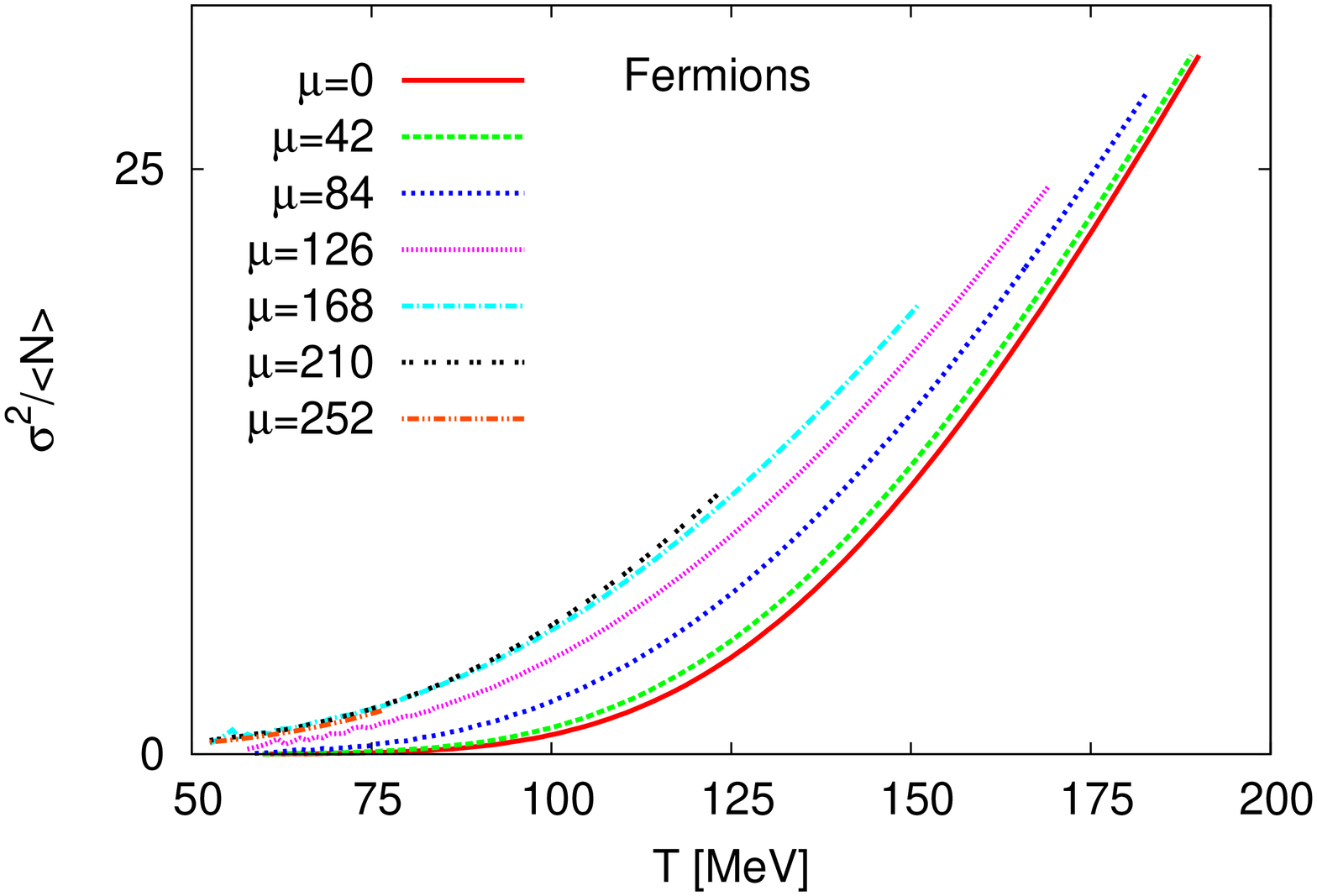}
\caption{The ratio $\sigma^2/\langle N\rangle$  is given in dependence on $T$ at different chemical potentials $\mu$ for hadronic (left), bosonic (middle) and fermionic (right) resonance gas. }
\label{fig:S2n} 
\end{figure}

\begin{figure}[htb]
\includegraphics[angle=-90,width=8.cm]{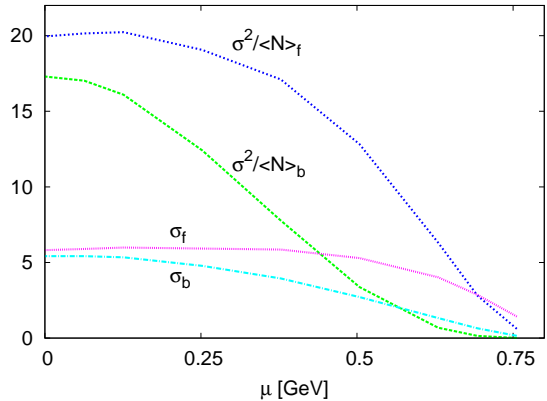}
\caption{The ratio $\sigma^2/\langle N\rangle$ and $\sigma$ are given in dependence on $\mu$ for boson and fermion resonances. }
\label{fig:S2n2} 
\end{figure}

The ratio of standard deviation $\sigma^2$ and the mean multiplicity $\langle N \rangle$ for fermions and bosons reads
\bea
\frac{\sigma^2}{\langle N\rangle} &=& \frac{1}{2}\, \frac{\int_0^{\infty}
  \left(1\pm \text{csch} \left[\frac{\varepsilon_i-\mu_i}{T}\right]\right)^{-1}\, k^2\, dk}{\int_0^{\infty} \left(1 \pm e^{\frac{\varepsilon_i-\mu_i }{T}}\right)^{-1}\, k^2\, dk} \label{eq:sMb}, 
\eea
where $\pm$ stands for fermions and bosons, respectively. The results are given in Fig. \ref{fig:S2n}. We notice that the bosonic resonance gas results in smaller values than the fermionic one, especially in final state. Furthermore, we notice that $\sigma^2/\langle N\rangle$ decreases with increasing $\mu$ of the bosons at middle temperature. This is exactly the opposite in the fermionic resonance gas. This would give an explanation for the observation that the results in hadron resonances are not spread as in the other two sectors. Figure \ref{fig:S2n2} shows the dependence of $\sigma^2/\langle N\rangle$ on $\mu$ at the chemical freeze-out boundary, which is characterized by $s/T^3=7$. The ratio $\sigma^2/\langle N\rangle$ is equivalent to $m_2/m_1$, Eq. (\ref{eq:gS2n}). The dependence of bosonic and fermionic $\sigma$ on $\mu$ is given in Fig. \ref{fig:S2n2}, as well. We notice that $\sigma$ is smaller than $\sigma^2/\langle N\rangle$, especially at small $\mu$. At large $\mu$ both quantities are almost equal. 

\begin{figure}[htb]
\includegraphics[angle=-90,width=5.5cm]{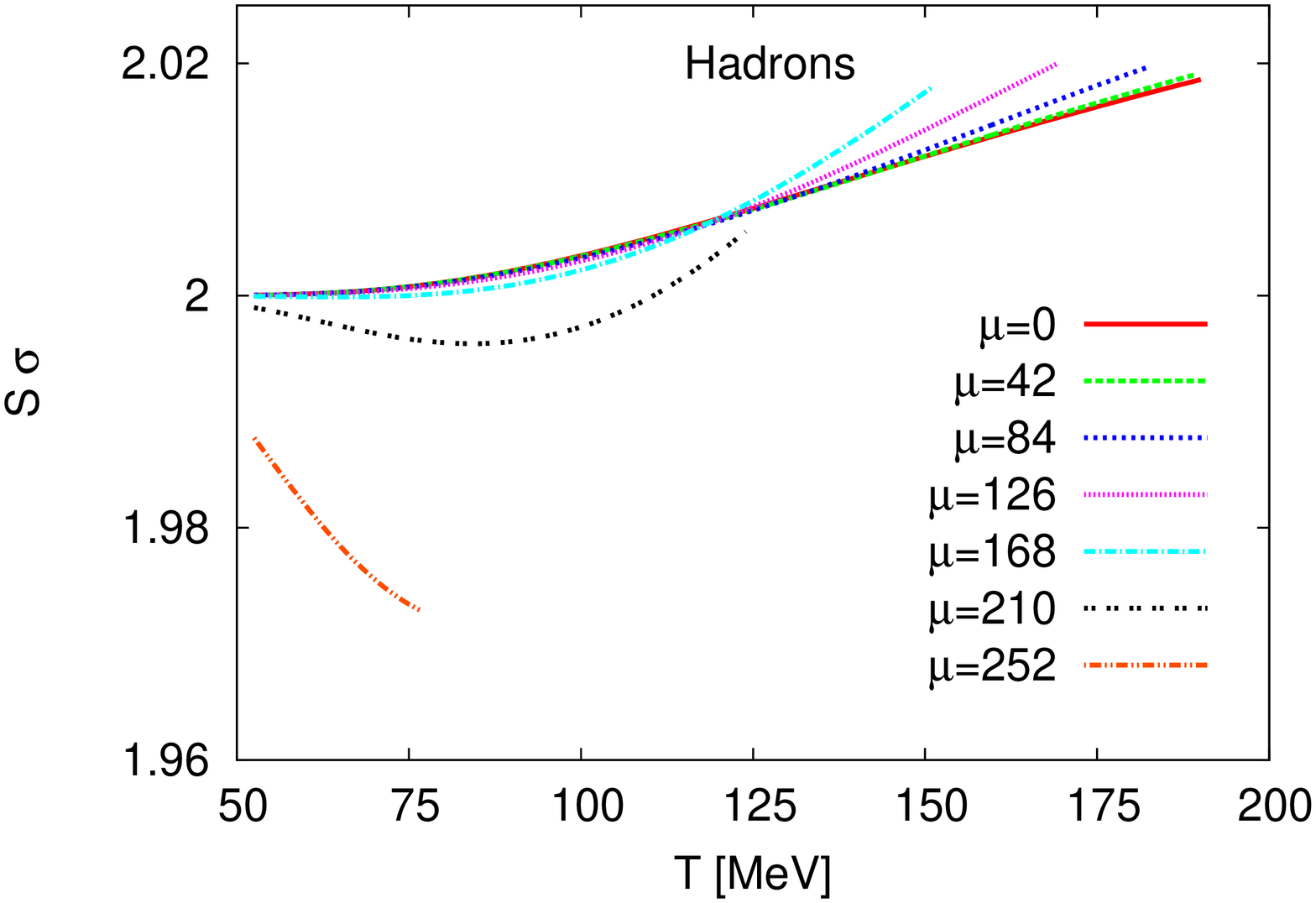}
\includegraphics[angle=-90,width=5.5cm]{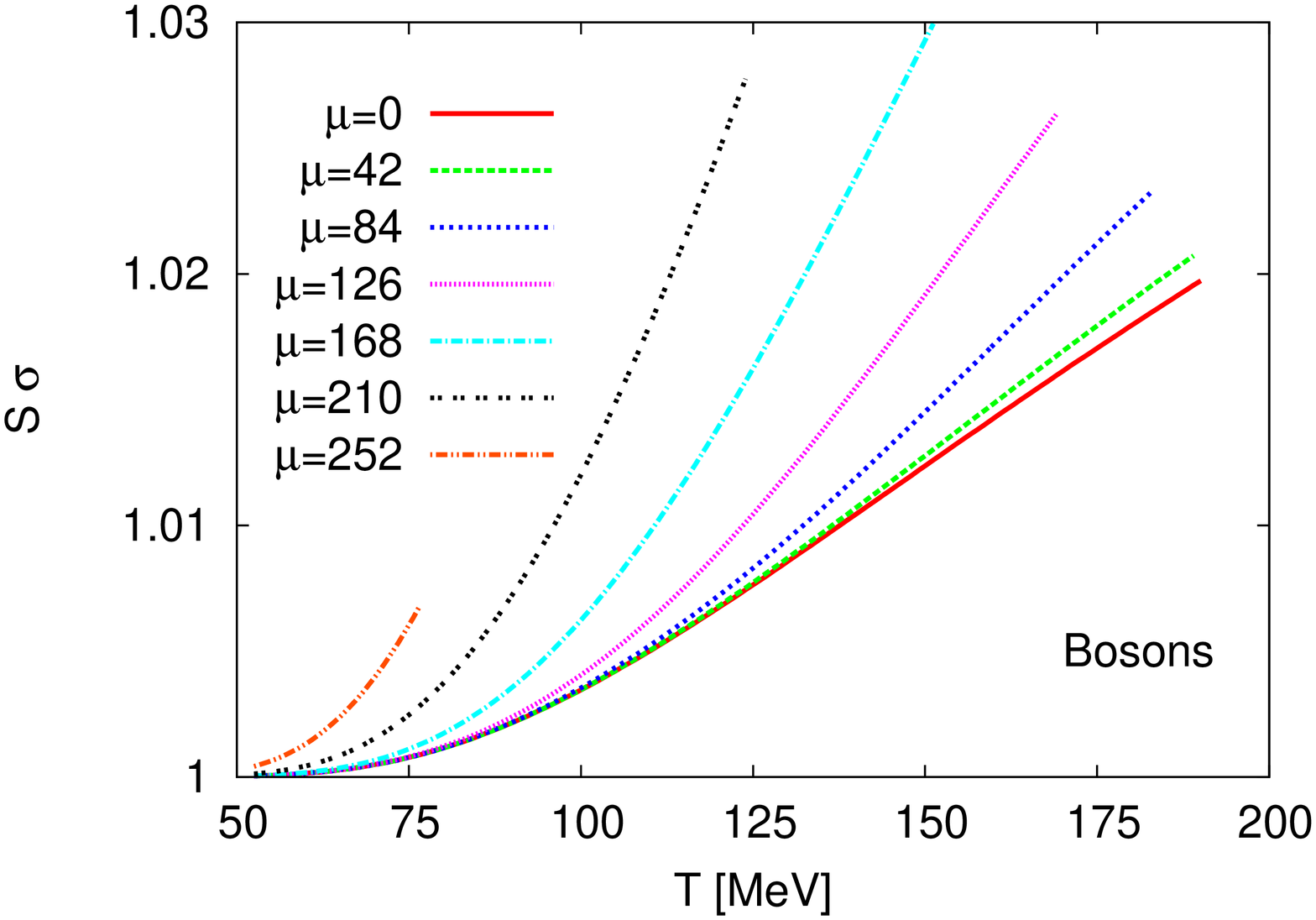}
\includegraphics[angle=-90,width=5.5cm]{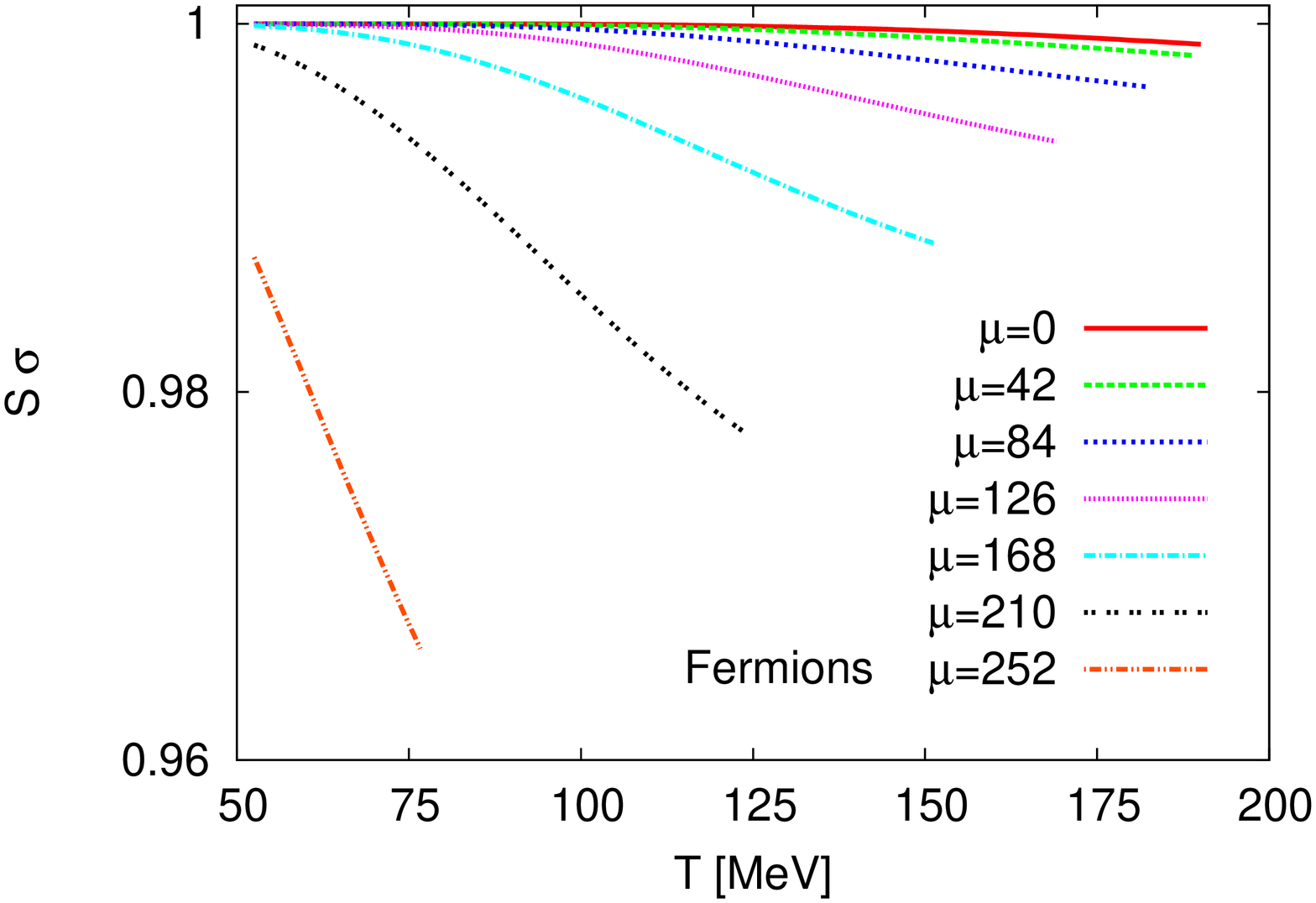}
\caption{The product $S \sigma$  is given in dependence on $T$ at different $\mu_b$ values (given in MeV).}
\label{fig:skewnS1} 
\end{figure}

The multiplication of skewness $S$ by the standard deviation $\sigma$ is directly related to the thermodynamics of the number susceptibility of the lattice QCD. In HRG, the bosonic and fermionic products read
\bea 
\left(S\; \sigma\right)_b &=& -\frac{1}{4}\; \frac{\int_0^{\infty} \, 
\text{csch}\left[\frac{\varepsilon_i -\mu_i }{T}\right]^4\; \text{sinh}\left[\frac{\varepsilon_i
      -\mu_i }{T}\right]\; k^2\,dk}{\int_0^{\infty} \left(1-\text{cosh}
  \left[\frac{\varepsilon_i -\mu_i }{T}\right]\right)^{-1}\; k^2\,dk}, \label{eq:ssigma1b} \\
\left(S\; \sigma\right)_f &=&  4 \; \frac{\int_0^{\infty} \, \text{csch}\left[\frac{\varepsilon_i
      -\mu_i }{T}\right]^3\; \text{Sinh}\left[\frac{\varepsilon_i
      -\mu_i }{2\, T}\right]^4\; k^2\,dk}{\int_0^{\infty} \left(\text{cosh}
  \left[\frac{\varepsilon_i -\mu_i }{T}\right]+1\right)^{-1}\; k^2\,dk}. \label{eq:ssigma1f}
\eea
The product $S\, \sigma$ is equivalent to $m_3/m_2$, Eq. (\ref{eq:gSs2}). The results are given in Fig. \ref{fig:skewnS1}. It is obvious that $S\, \sigma \simeq 1$ for either bosons or fermions. Then, for hadrons, $S\, \sigma \simeq 2$. Nevertheless, the fine structure seems to reveal interesting features.
  
The dependence of $\sigma$ and $\sigma^2/\langle N\rangle$ on $T$ is illustrated in Figs. \ref{fig:sigmaaa} and \ref{fig:S2n}, respectively. It is obvious that both quantities have a monotonic behavior. Their dependences on $\mu$ are given in Fig. \ref{fig:S2n2}. Also, this type of dependences seems to be monotonic. As given in Fig. \ref{fig:skewnS1}, the product $S\, \sigma$ has a characteristic dependence on $T$. Regardless, the tiny change, we notice that increasing $\mu$ increases the bosonic $S\, \sigma$ product, but decreases the fermionic $S\, \sigma$ product. In both cases, it forms a folding fan. The hadronic product makes an amazing bundle in the middle (at a characteristic $T$). This will be discussed in details in section \ref{sec:chemFO}. When $S$ is studied as a function of $\mu$ and given in Fig. \ref{fig:kSmuu}, we find that $S(\mu)_b$ remains almost constant, while $S(\mu)_f$ raises with increasing $\mu$. We also notice that both curves cross at a certain point.  This is not the case of the $\mu$-dependence of $\sigma$ and $\sigma^2/\langle N\rangle$, Fig. \ref{fig:S2n2}. So far, we conclude that starting from the third normalized moment, a non-monotonic behavior appears. Thus, the kurtosis and its products should be expected to highlight such a non-monotonic behavior \cite{ritter11,endp5,qcdlike}. 

\begin{figure}[htb]
\includegraphics[angle=-90,width=5.5cm]{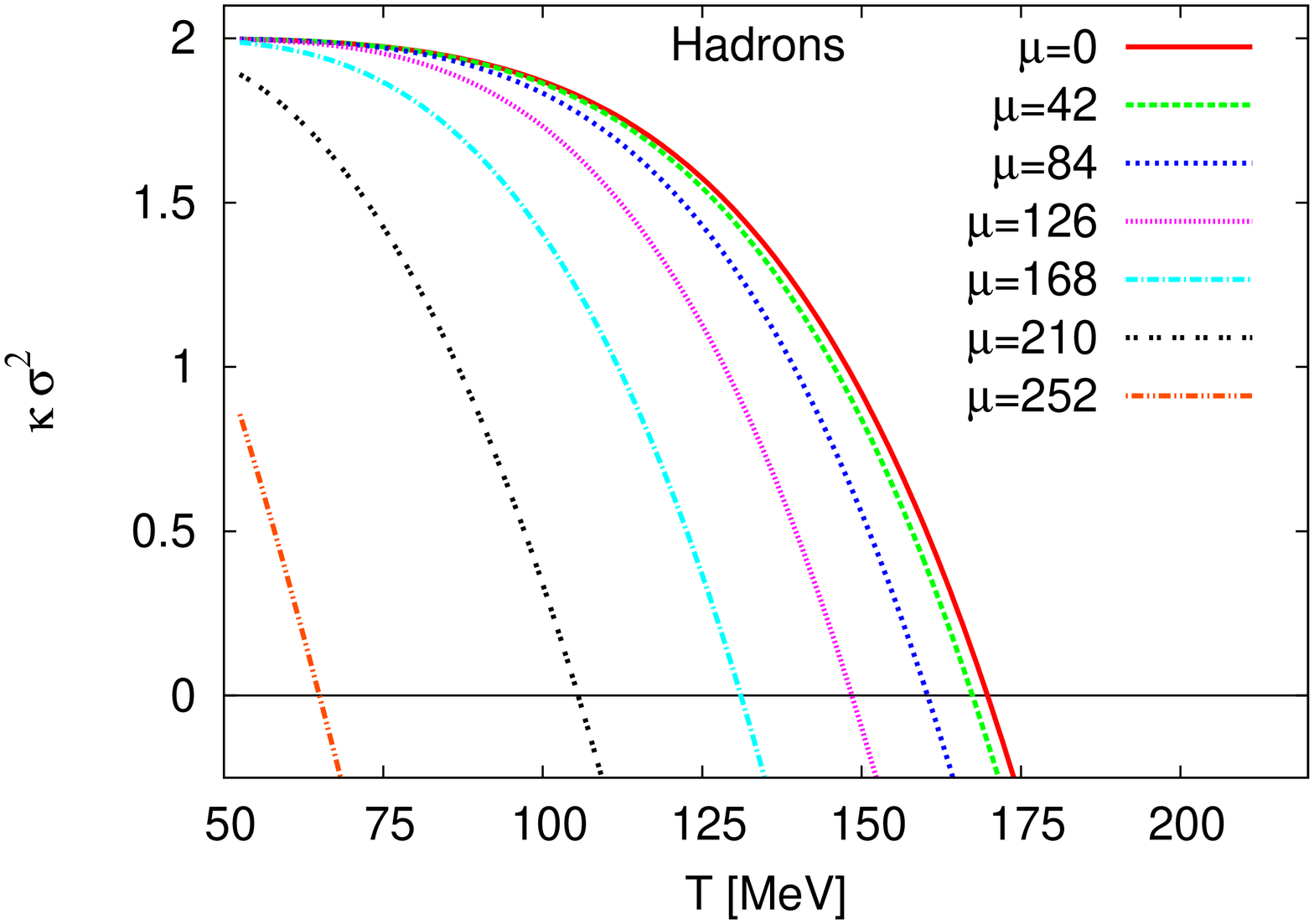}
\includegraphics[angle=-90,width=5.5cm]{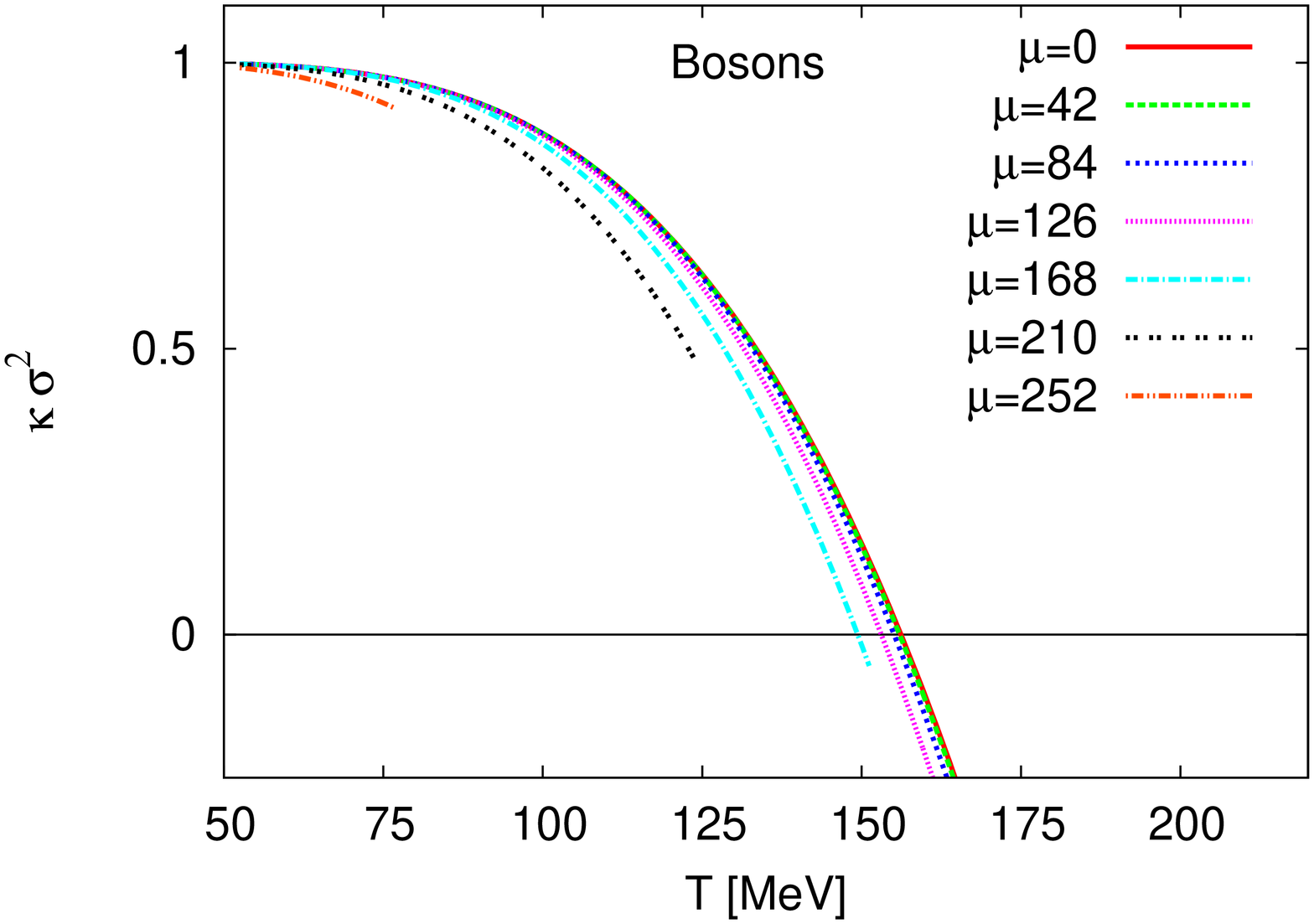}
\includegraphics[angle=-90,width=5.5cm]{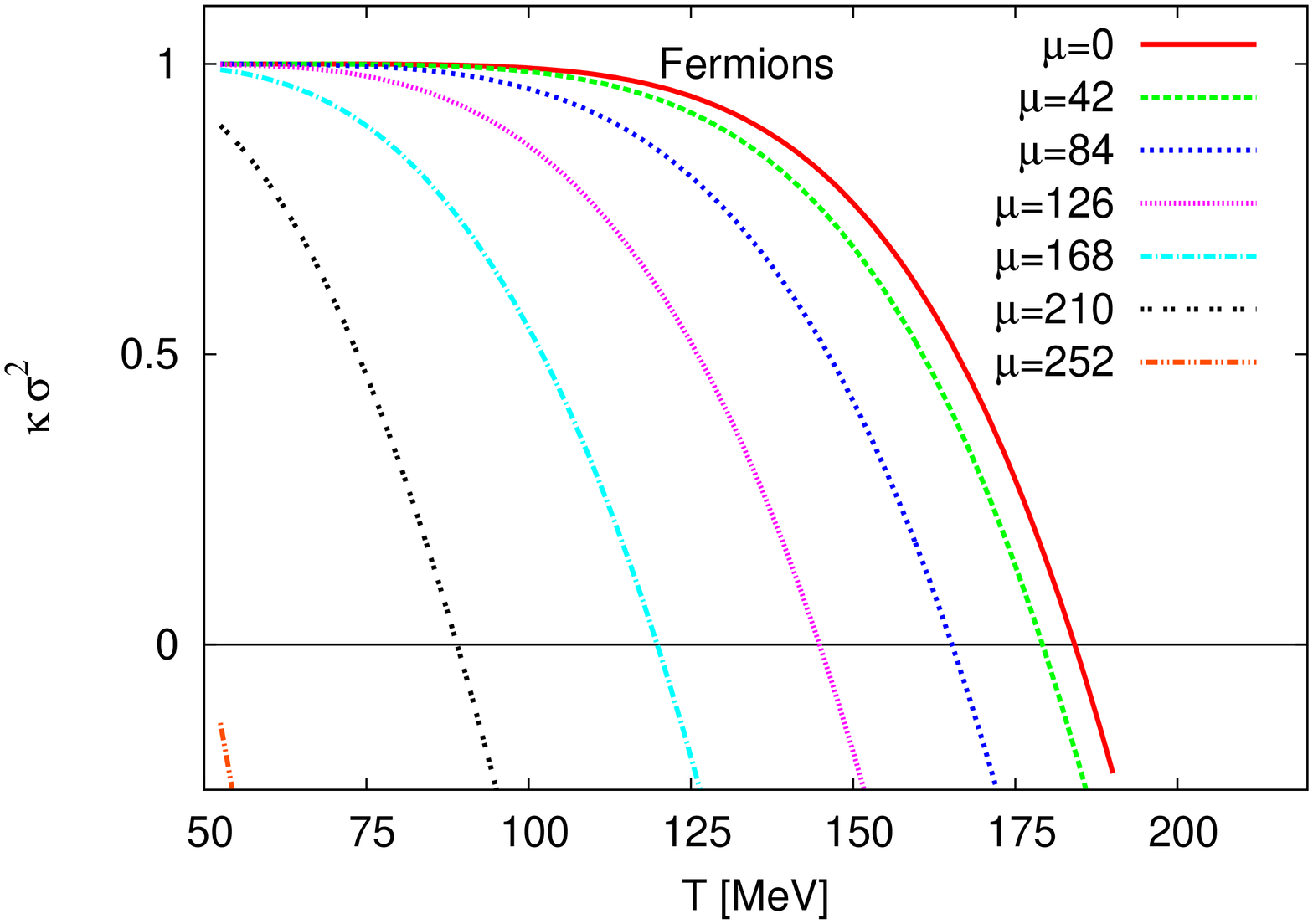}
\caption{The product ${\kappa}\,\sigma^2$ is given as a function of $T$ at various $\mu$-values for hadronic (left), bosonic (middle) and fermionic resonance gas (right).}
\label{fig:kS2muu} 
\end{figure}

\begin{figure}[htb]
\includegraphics[angle=-90,width=8.cm]{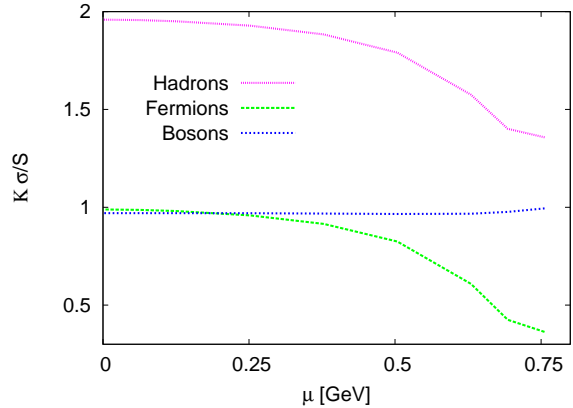}
\caption{The dependence of ${\kappa}\,\sigma/S$ on the chemical potential $\mu$.}
\label{fig:kSsmuub} 
\end{figure}

The multiplication of kurtosis by $\sigma^2$ called $\kappa^{eff}$ \cite{keff} is apparently equivalent to the ratio of $3$-rd order moment to $2$-nd order moment, Eq. (\ref{eq:gks2}).  In lattice QCD  and QCD-like models, $\kappa^{eff}$ is found to diverge near the critical endpoint \cite{endp5,qcdlike}. In HRG, the bosonic and fermionic products read
\bea 
\left(\kappa\; \sigma^2\right)_b &=& -\frac{1}{4} \frac{\int_0^{\infty}  \left\{\text{cosh}\left[\frac{\varepsilon_i
      -\mu_i }{T}\right] + 2\right\}\; \text{csch}\left[\frac{\varepsilon_i
      -\mu_i }{2\, T}\right]^4 \; k^2\,dk} {\int_0^{\infty} \left(1-\text{cosh}
  \left[\frac{\varepsilon_i -\mu_i }{T}\right]\right)^{-1}\; k^2\,dk} \nonumber \\
&+& \frac{3\, g_i}{4\, \pi^2}\, \frac{1}{T^3}\, \int_0^{\infty} \left(1-\text{cosh}
  \left[\frac{\varepsilon_i -\mu_i }{T}\right]\right)^{-1}\; k^2\,dk, \hspace*{7mm}\label{eq:lsigma2b} \\
\left(\kappa\; \sigma^2\right)_f &=& \frac{1}{4} \frac{\int_0^{\infty}  \left\{\text{cosh}\left[\frac{\varepsilon_i
      -\mu_i }{T}\right] - 2\right\}\; \text{Sech}\left[\frac{\varepsilon_i
      -\mu_i }{2\, T}\right]^4 \; k^2\,dk} {\int_0^{\infty} \left(\text{cosh}
  \left[\frac{\varepsilon_i -\mu_i }{T}\right]+1\right)^{-1}\; k^2\,dk} \nonumber \\
&-& \frac{3\, g_i}{4\, \pi^2}\, \frac{1}{T^3}\, \int_0^{\infty} \left(\text{cosh}
  \left[\frac{\varepsilon_i -\mu_i }{T}\right]+1\right)^{-1}\; k^2\,dk. \label{eq:lsigma2f}
\eea
When ignoring the constant term in Eqs. (\ref{eq:Kkb}) and (\ref{eq:Kkf}), then the second terms in the previous expressions entirely disappear. The results are given in Fig. \ref{fig:kS2muu}. In the hadronic sector, the dependence of $\kappa\,\sigma^2$ on the temperature $T$ at different $\mu$-values is given in the left panel. We notice that increasing $T$ is accompanied with a drastic declination in $\kappa\,\sigma^2$. Also, we find that ${\kappa}\,\sigma^2$ flips its sign at large $T$. When comparing the thermal evolution of $\kappa\,\sigma^2$ with the one of $S\, \sigma$, Fig. \ref{fig:skewnS1}, we simply find that the latter is much drastically changing than the earlier. Also, when comparing their dependences on the chemical potentials at the freeze-out boundary, Figs. \ref{fig:kSsmuub} and \ref{fig:fezeout-sT3}, it is apparent that the $\mu$-dependence increases when the normalized fourth order moment is included. The product $\kappa\, \sigma^2$ calculated at the freeze-out boundary leads to some interesting findings. First, $\kappa\, \sigma^2$ almost vanishes or even flips its sign. Second, the $T$ and $\mu$ corresponding to vanishing $\kappa\, \sigma^2$ are coincident with the phenomenologically measured freeze-out parameters. Third, the freeze-out boundaries of bosons and fermions are crossing at a point located very near to the one assumed by the lattice QCD calculations to be the QCD CEP, section \ref{sec:cep}.

\begin{figure}[htb]
\includegraphics[angle=-90,width=5.5cm]{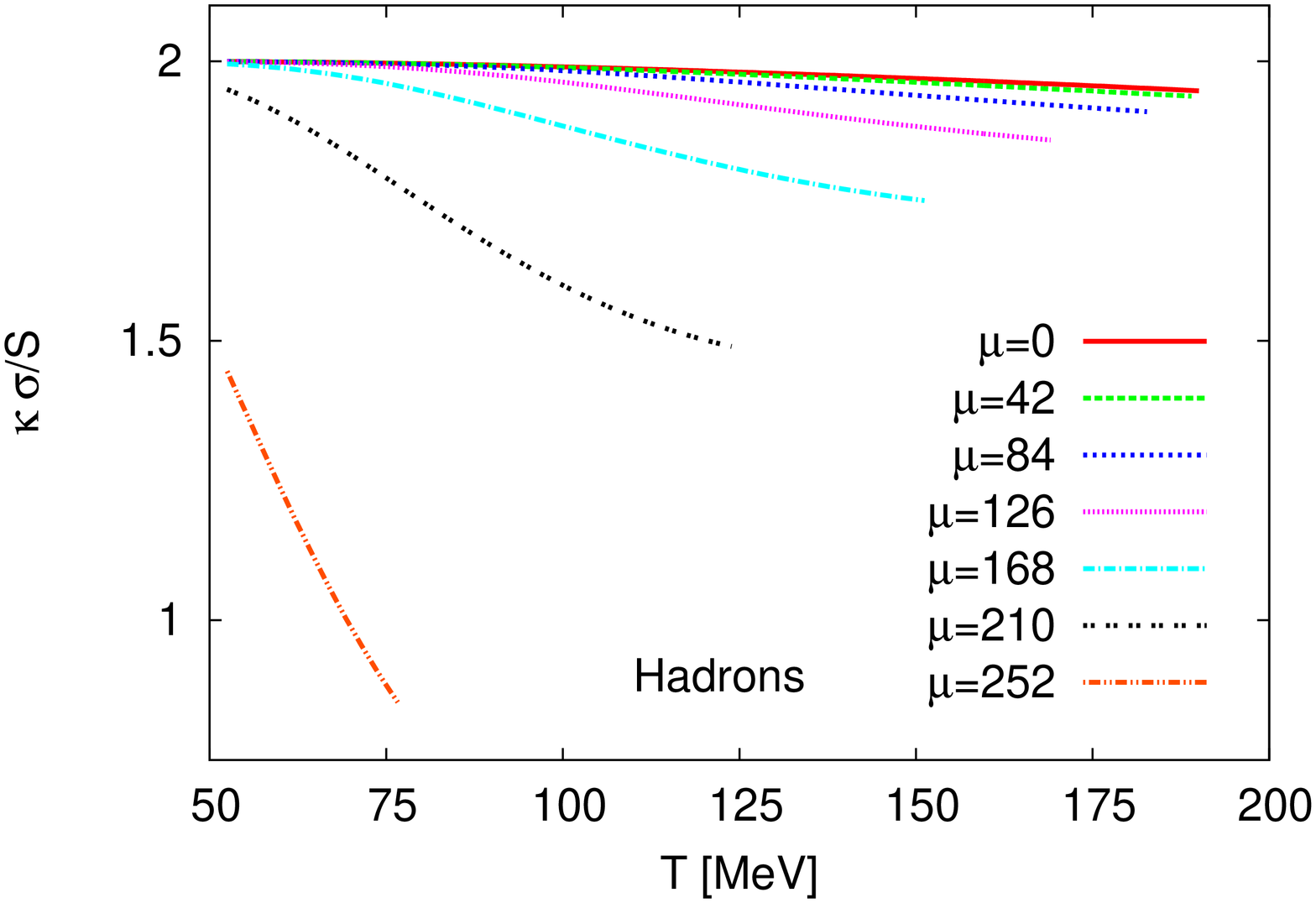}
\includegraphics[angle=-90,width=5.5cm]{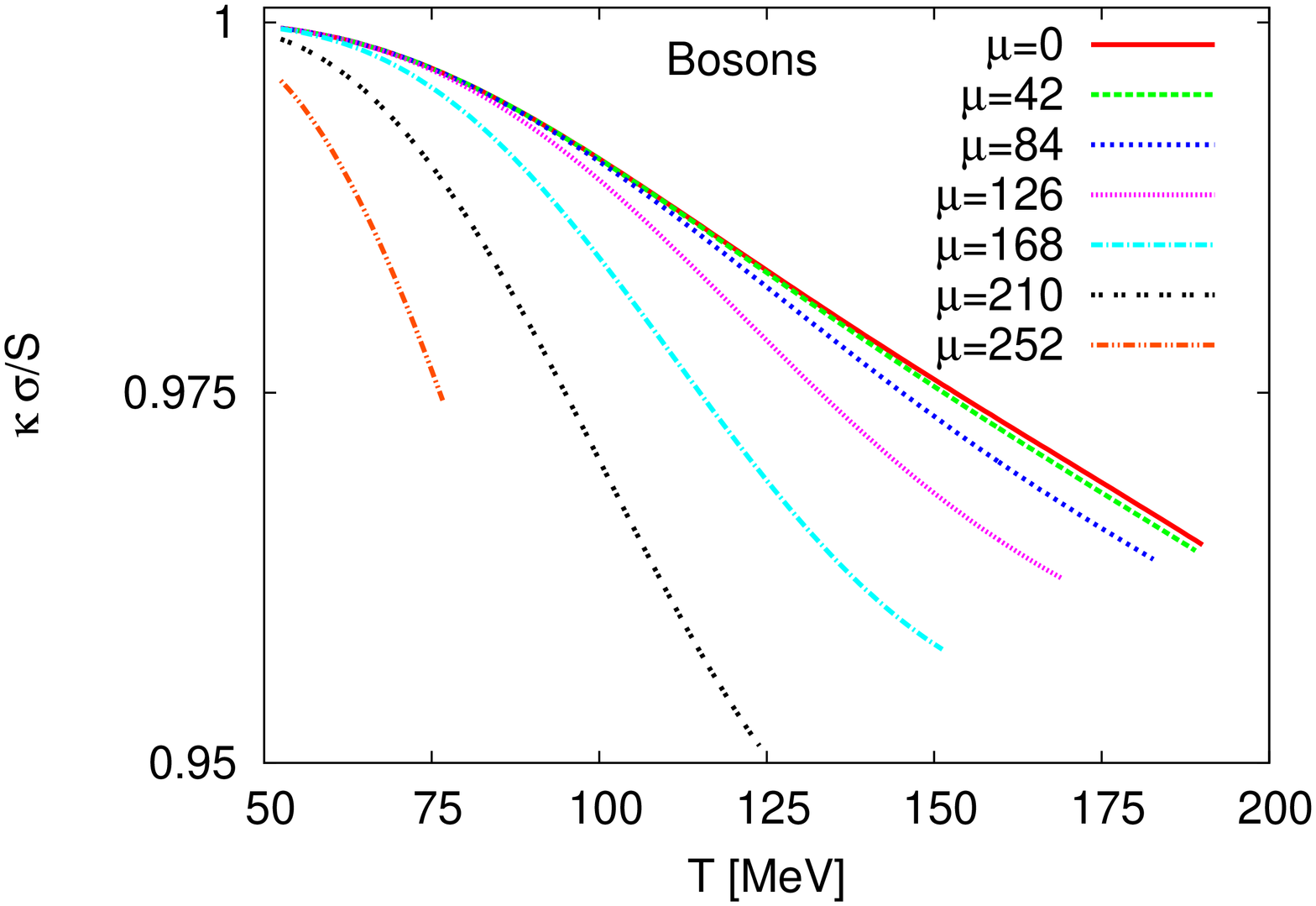}
\includegraphics[angle=-90,width=5.5cm]{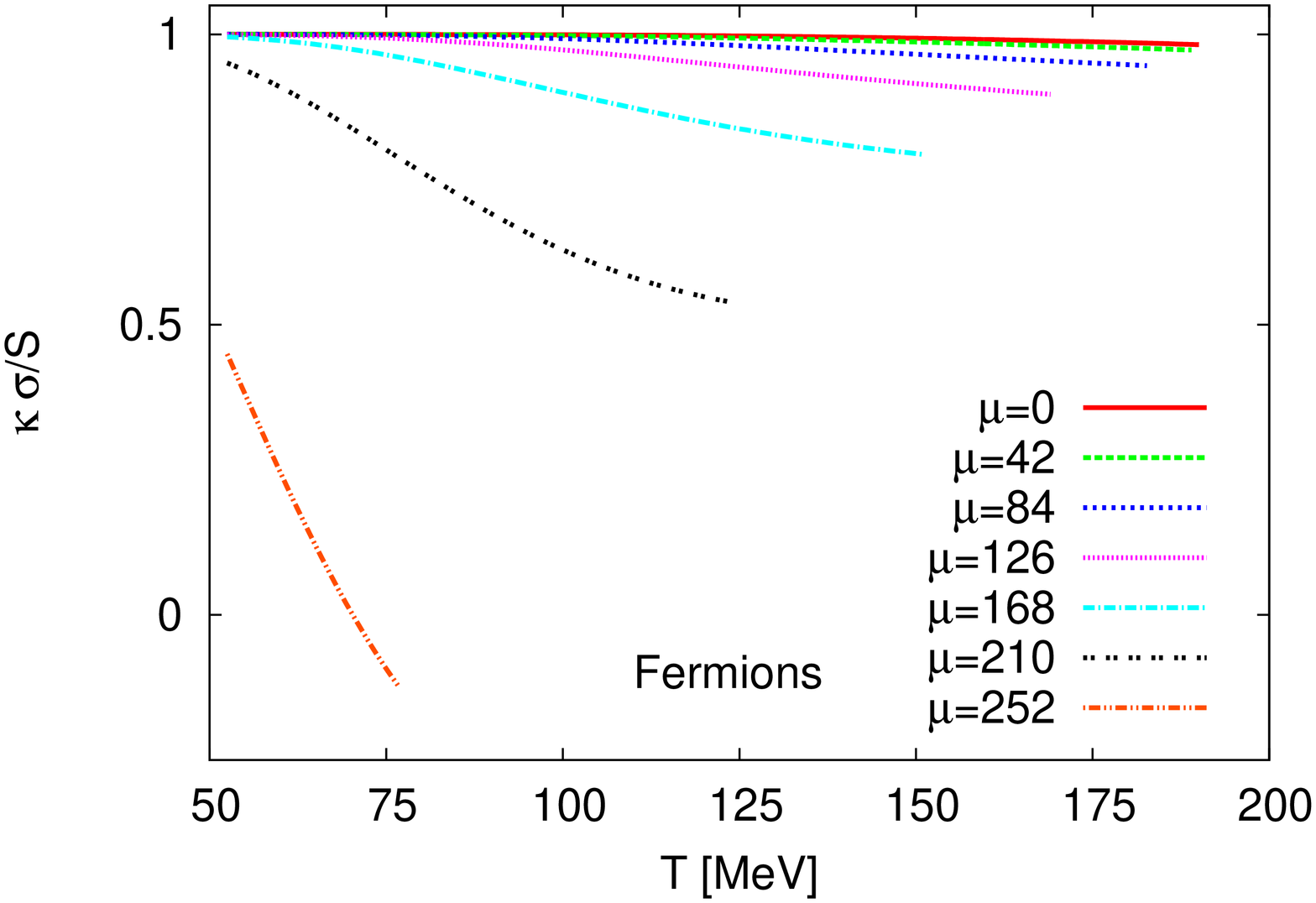}
\caption{${\kappa}\,\sigma/S$ as a function $T$ at different chemical potentials for hadrons (left), bosons (middle) and fermions (right).}
\label{fig:kSsmuu} 
\end{figure}

The skewness and kurtosis can be combined through the standard deviation as follows.
\bea 
\left(\frac{\kappa\; \sigma}{S}\right)_b &=& 
\frac{-\frac{1}{4}\,\pi^2\, T^3\,
\int_0^{\infty} \left\{2 + \text{cosh}\left[\frac{\varepsilon_i -\mu_i
  }{T}\right]\right\} \text{csch} \left[\frac{\varepsilon_i -\mu_i }{2\,
    T}\right]^4 \; k^2\, dk + \frac{3}{4}\, g_i\, \left(\int_0^{\infty}
  \left(1-\text{cosh}\left[\frac{\varepsilon_i -\mu_i }{T}\right]\right)^{-1}\; k^2\, dk\right)^2 }
{\frac{1}{4}\, \pi^2\, T^3\, \int_0^{\infty} \text{csch}
  \left[\frac{\varepsilon_i -\mu_i }{2\, T}\right]^4\; \text{sinh}
  \left[\frac{\varepsilon_i -\mu_i }{T}\right]\; k^2\,dk }, \label{ksigmaSb} \\
\left(\frac{\kappa\; \sigma}{S}\right)_f &=& 
\frac{\frac{1}{4}\,\pi^2\, T^3\, 
\int_0^{\infty} \left\{\text{cosh}\left[\frac{\varepsilon_i -\mu_i
  }{T}\right]-2 \right\} \text{sech} \left[\frac{\varepsilon_i -\mu_i }{2\,
    T}\right]^4 \; k^2\, dk - \frac{3}{4}\, g_i\, \left(\int_0^{\infty}
  \left(1+\text{cosh}\left[\frac{\varepsilon_i -\mu_i }{T}\right]\right)^{-1}\; k^2\, dk \right)^2 }
{4\, \pi^2\,T^3\, \int_0^{\infty} \text{csch}
  \left[\frac{\varepsilon_i -\mu_i }{T}\right]^3\; \text{sinh}
  \left[\frac{\varepsilon_i -\mu_i }{2\, T}\right]^4\; k^2\,dk }. \label{ksigmaSf} \hspace*{10mm} 
\eea
Also when ignoring the constant term in Eqs. (\ref{eq:Kkb}) and (\ref{eq:Kkf}),
the second terms in previous expressions disappear. The results of $\kappa\, \sigma/S$ are given in Fig. \ref{fig:kSsmuu}. In the fermionic sector, $\kappa\, \sigma/S$ decreases with increasing $T$. Increasing $\mu$ makes the decrease much faster. Although the drastic change in $\kappa\, \sigma/S$, this behavior can be compared with $S\, \sigma$, Fig. \ref{fig:skewnS1}, qualitatively. In the bosonic sector, there is a very slow decrease with increasing $\mu$. The same sector in $S\, \sigma$, Fig. \ref{fig:skewnS1}, shows a very slow but an increasing dependence on $T$. Therefore,  in the hadronic sector, $\kappa\, \sigma/S$ is overall diminishing with increasing $T$. The declination becomes faster for larger $\mu$.  Fig. \ref{fig:kSsmuub} illustrates the $\mu$-dependence, i.e. $\kappa\, \sigma/S$ is estimated at the freeze-out curve. We notice that the bosons result in an almost unvarying $\kappa\, \sigma/S$ with growing $\mu$. Therefore, the overall decrease in the hadronic sector is originated in the fermionic degrees of freedom.

From Eqs. (\ref{eq:sMb})-(\ref{ksigmaSf}), we summarize that  
\bea
\frac{\sigma^2}{\langle N\rangle} & \equiv & \frac{m_2}{m_1}, \label{eq:gS2n} \\
S\; \sigma & \equiv & \frac{m_3}{m_2}, \label{eq:gSs2}\\
\kappa\; \sigma^2 & \equiv & \frac{m_4}{m_2}, \label{eq:gks2}\\
\frac{\kappa\; \sigma}{S} & \equiv & \frac{m_4}{m_3}. \label{eq:gksS}
\eea
Obviously, other products would complete missing ratios, for example,
\bea
S\, \chi  &\equiv & \frac{m_3}{m_1}, \\
\frac{\kappa\, \chi^2}{\langle N\rangle}  &\equiv & \frac{m_4}{m_1},
\eea
relating third and fourth order moments to the first one, respectively. As discussed above, the susceptibility $\chi$ is equivalent to $\sigma^2$. These products seem to be sensitive to the volume independent multiplicities.

\begin{figure}[htb]
\includegraphics[angle=-90,width=5.5cm]{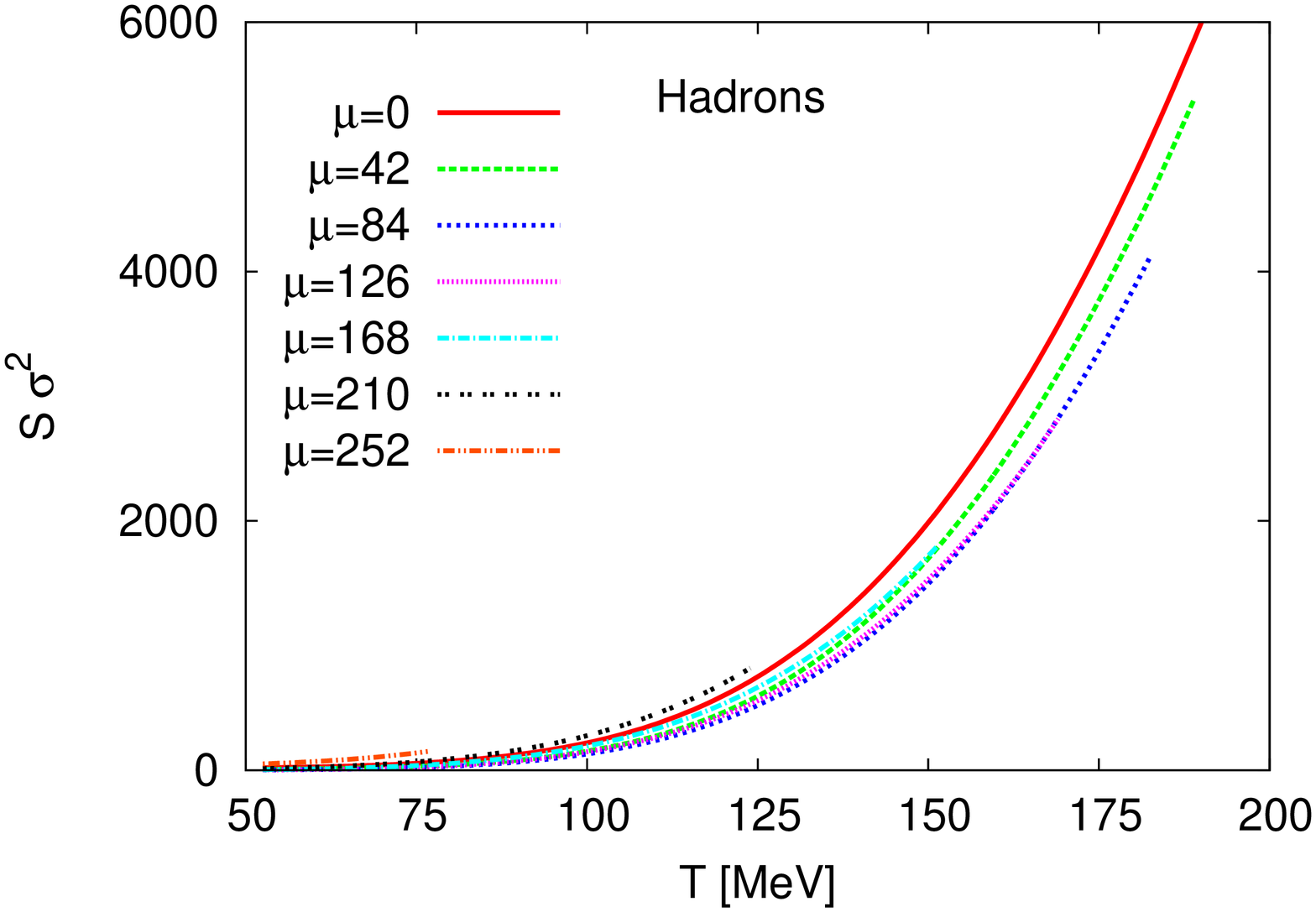}
\includegraphics[angle=-90,width=5.5cm]{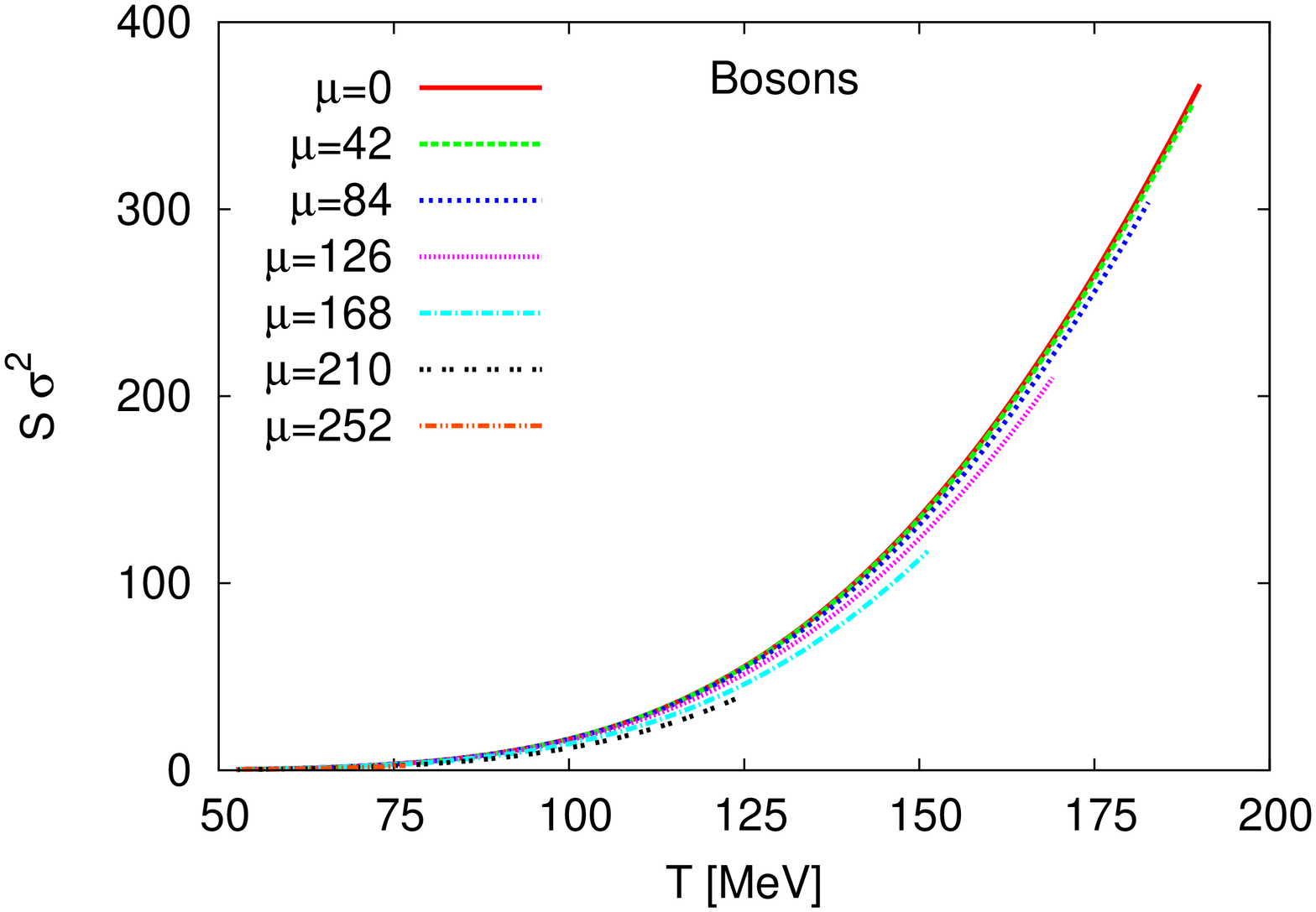}
\includegraphics[angle=-90,width=5.5cm]{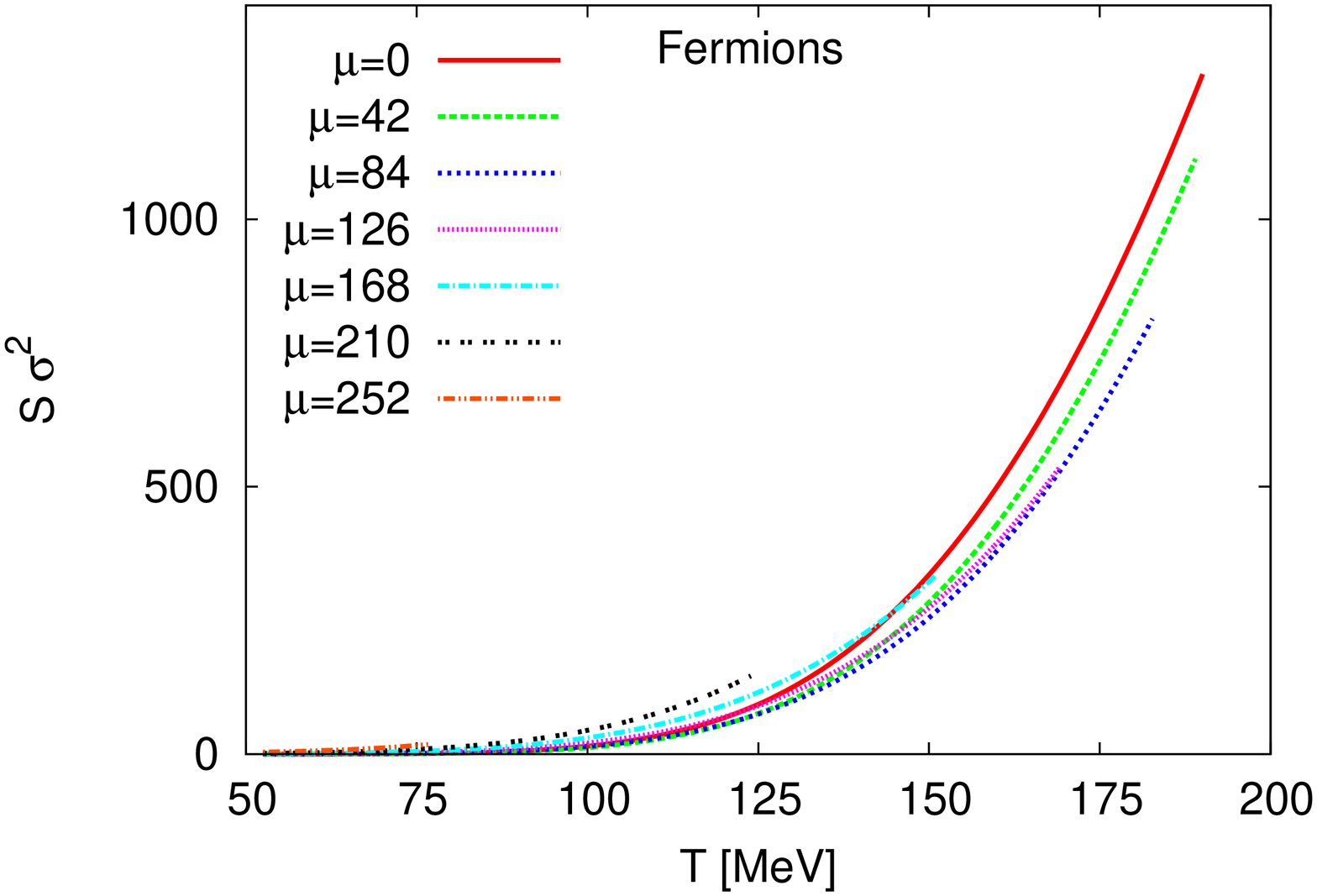}
\caption{The thermal evolution of $S\, \sigma^2$ for hadrons (left), bosons (middle) and fermions (right) at different $\mu$-values (given in MeV).}
\label{fig:Schi1} 
\end{figure}

\begin{figure}[htb]
\includegraphics[angle=-90,width=8.cm]{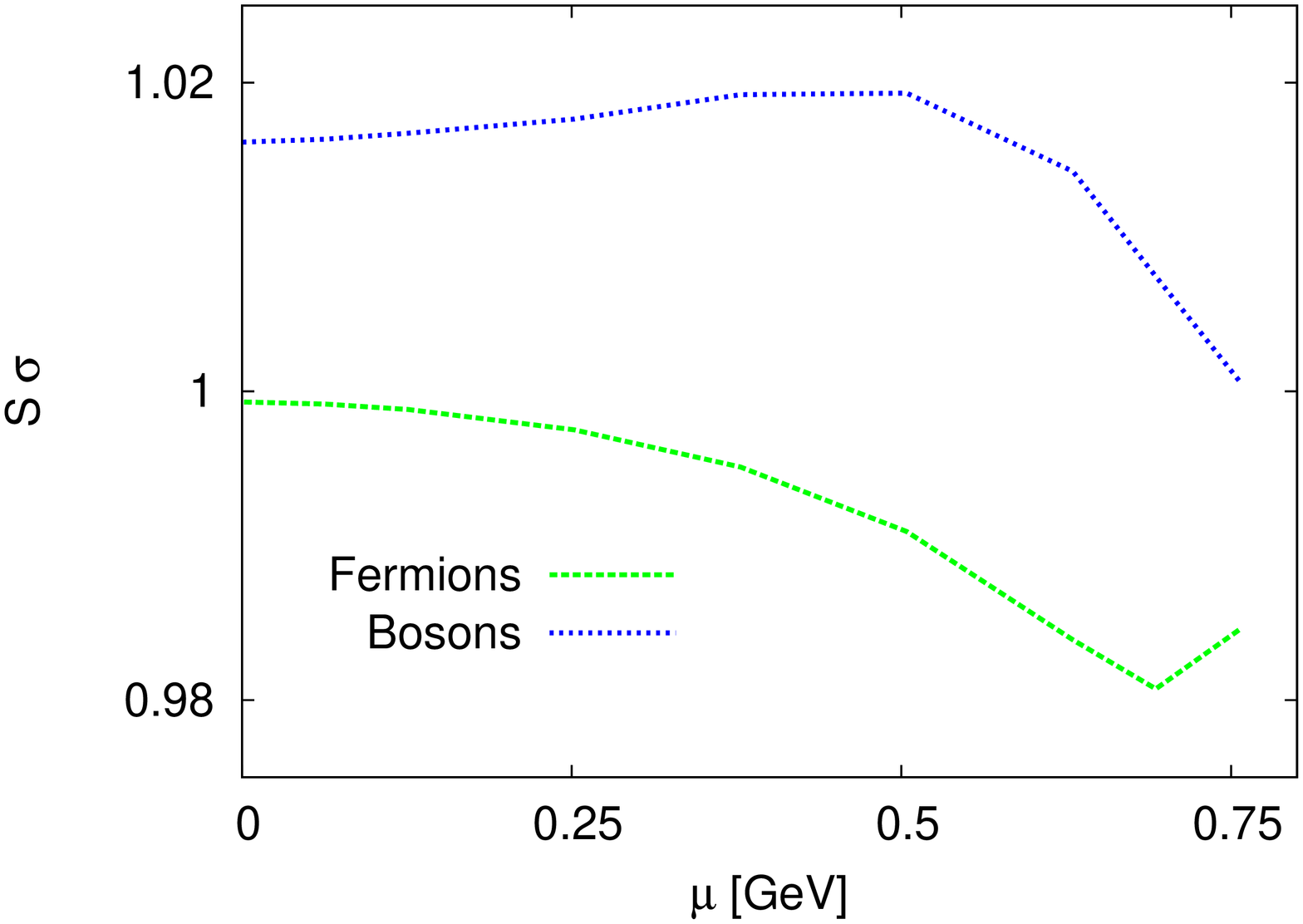}
\includegraphics[angle=-90,width=8.cm]{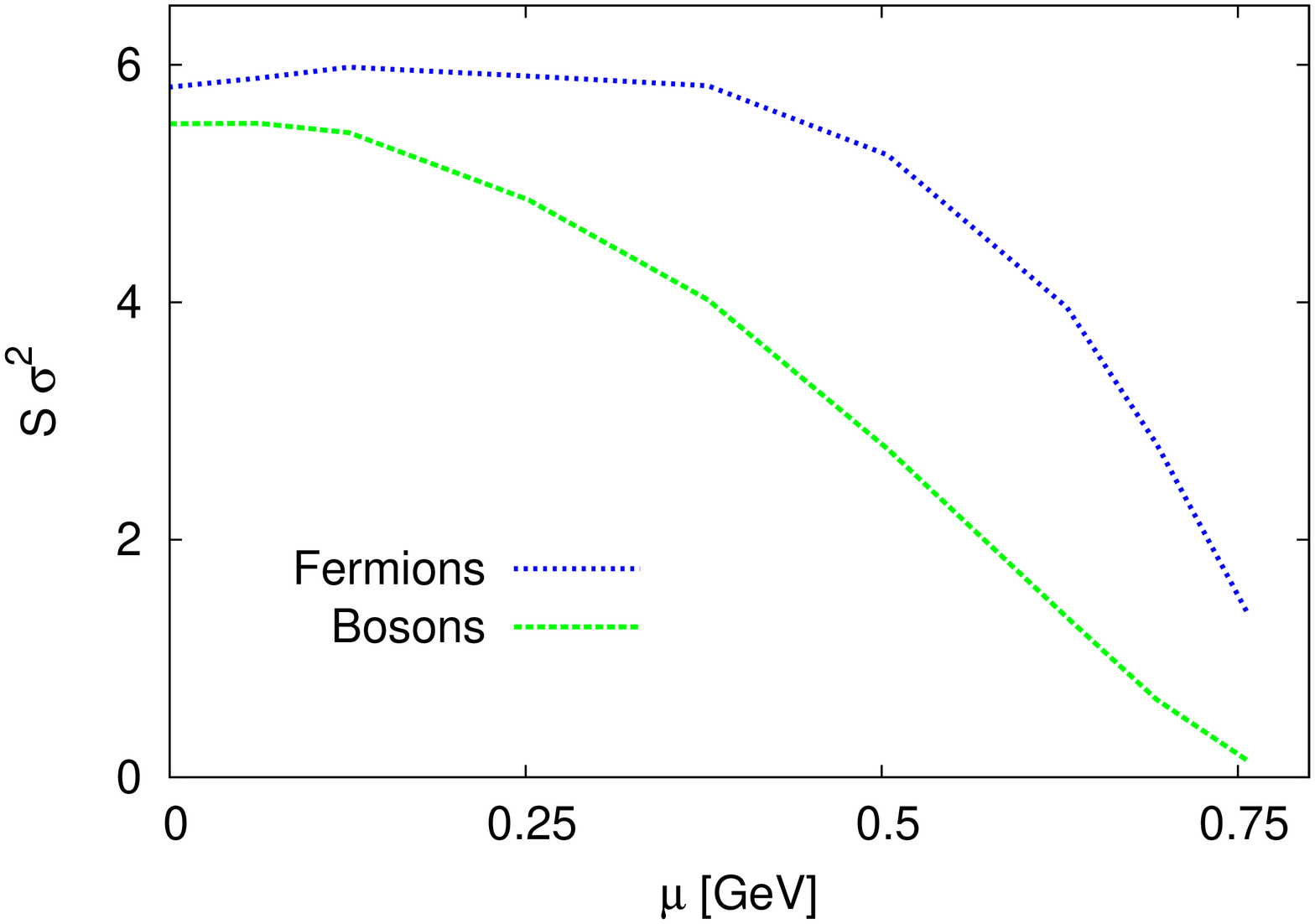}
\caption{The dependence of $S\, \sigma$ (left panel) and $S\,\sigma^2$ (right panel) of bosons and fermions on the chemical potential $\mu$ (given in MeV).}
\label{fig:Schi2} 
\end{figure}

The results of $S\,\chi\equiv S\,\sigma^2$ are given in Fig. \ref{fig:Schi1}. At the freeze-out boundary, the $\mu$-dependence of $S\, \sigma^2$ seems to be much stronger than the $\mu$-dependence on $S\, \sigma$, Fig. \ref{fig:Schi2}.  

\begin{figure}[htb]
\includegraphics[angle=-90,width=5.5cm]{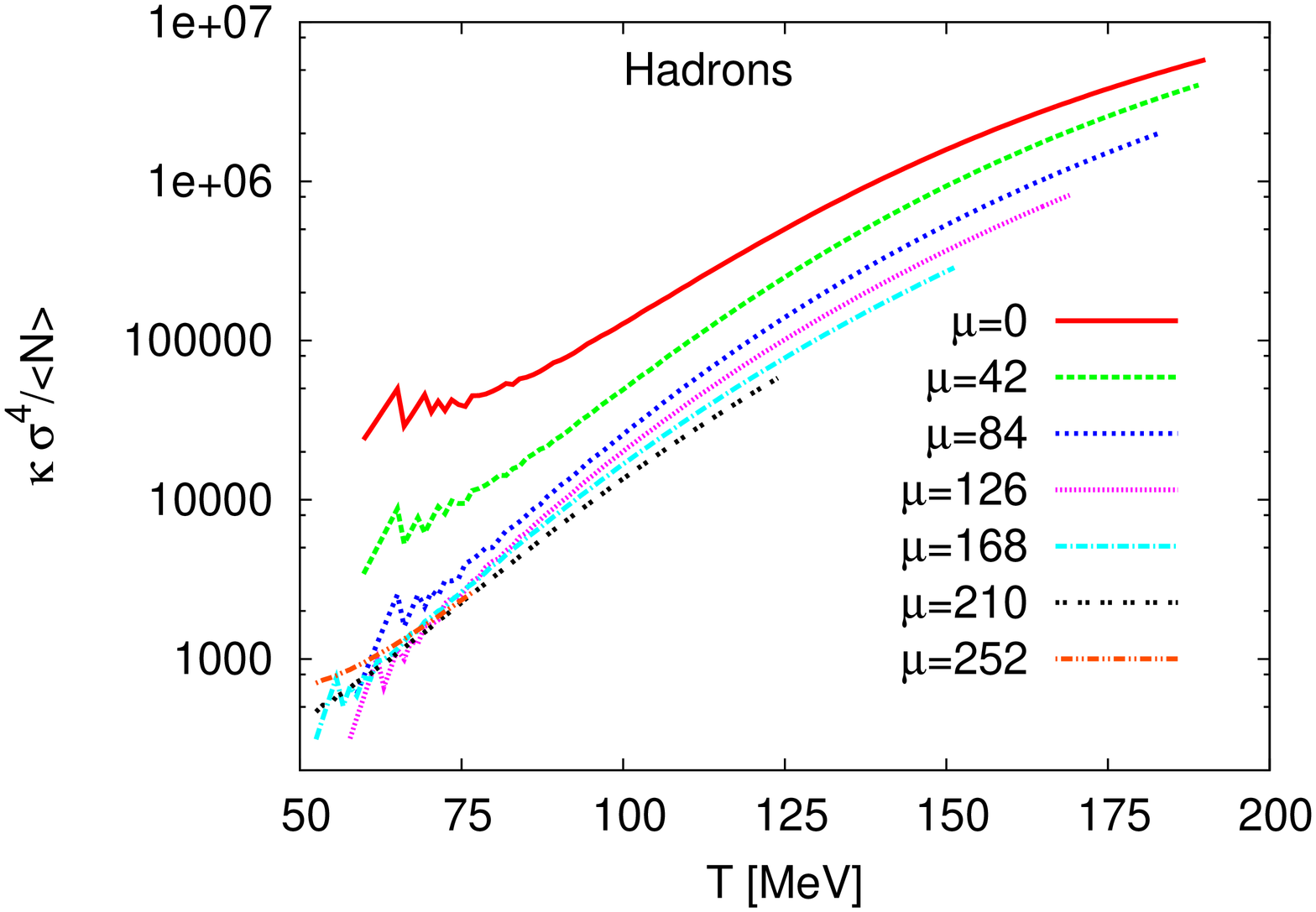}
\includegraphics[angle=-90,width=5.5cm]{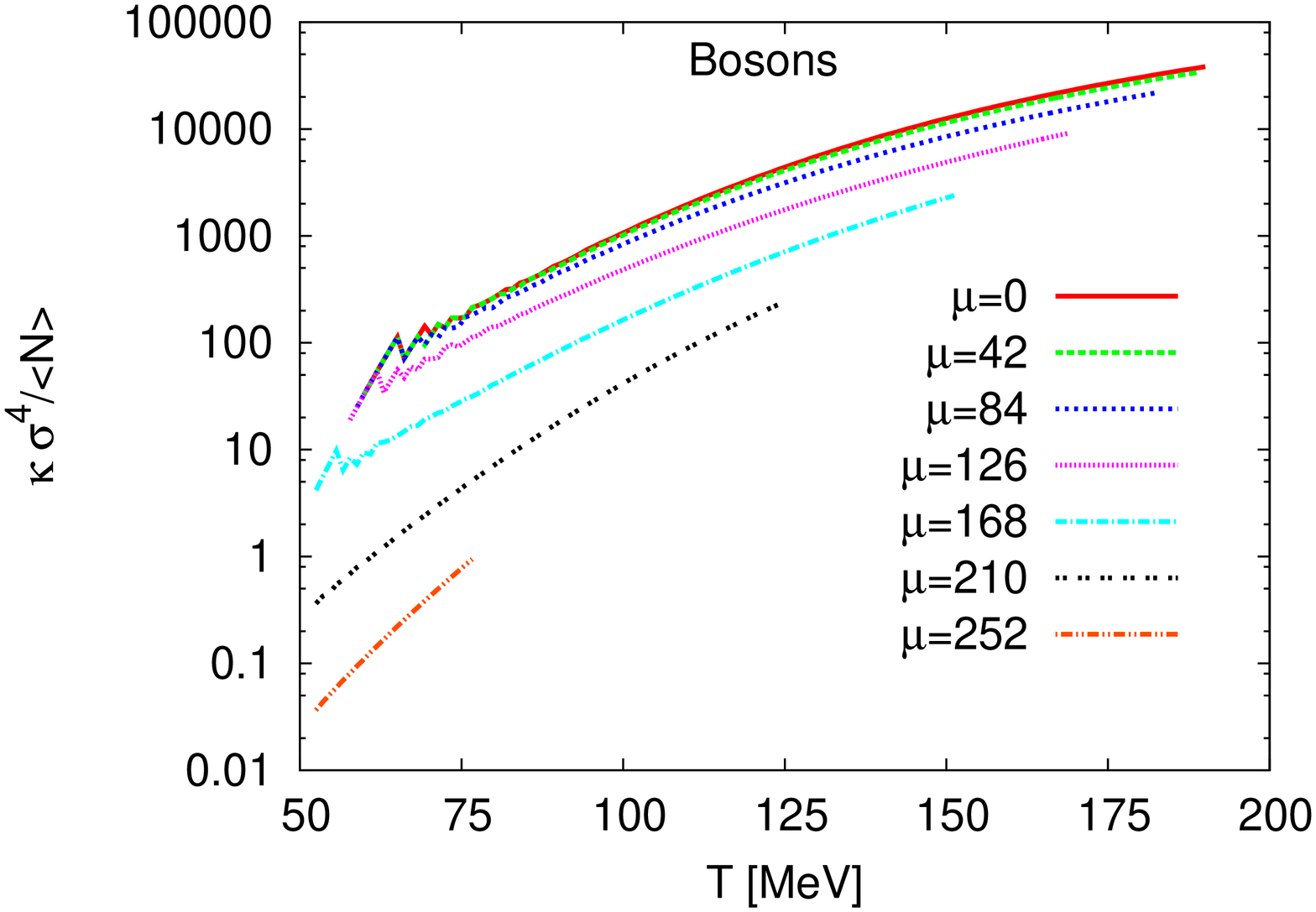}
\includegraphics[angle=-90,width=5.5cm]{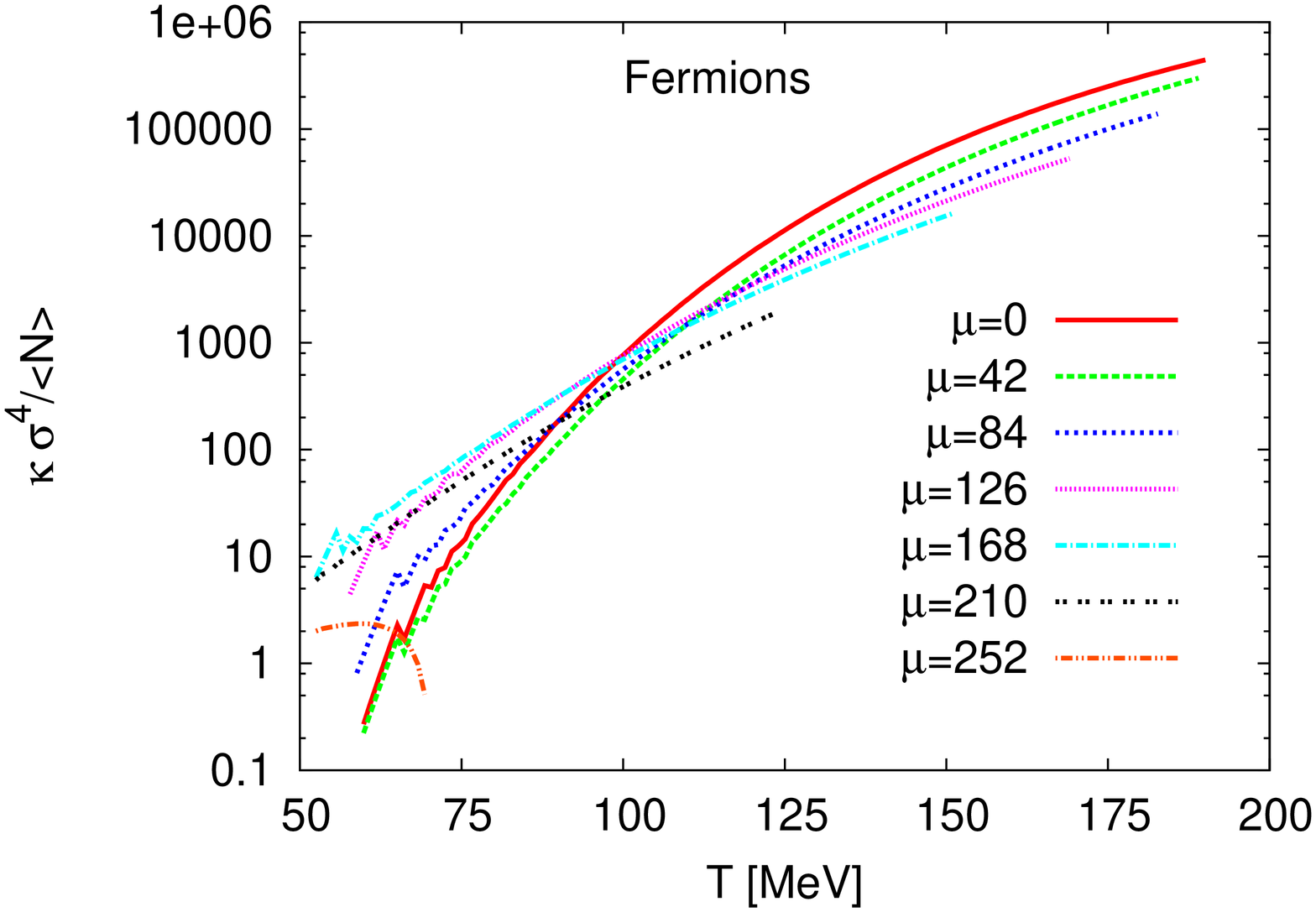}
\caption{Thermal evolution of $\kappa\,\sigma^4/\langle N\rangle$ for hadrons (left), bosons (middle) and fermions (right) at different chemical potentials (given in MeV).}
\label{fig:kS4nmuu1} 
\end{figure}

\begin{figure}[htb]
\includegraphics[angle=-90,width=8.cm]{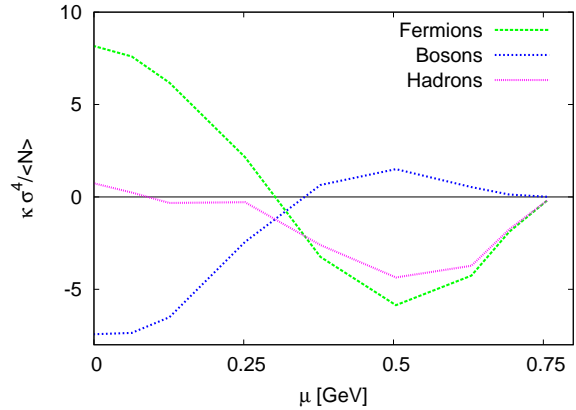}
\caption{The ratio $\kappa\, \sigma^4/\langle N\rangle$ are given in dependence on $\mu$ for boson and fermion resonances. }
\label{fig:kS4nsT3} 
\end{figure}
   
The thermal evolution of $\kappa\, \sigma^4/\langle N\rangle$ is illustrated in Fig. \ref{fig:kS4nmuu1}. Increasing $T$ is accompanied by an almost overall increase of  $\kappa\, \sigma^4/\langle N\rangle$ values. At low $T$, fluctuations appear. In bosonic sector, increasing $\mu$ leads to an obvious decrease in $\kappa\, \sigma^4/\langle N\rangle$ so that the curves are distributed over a wide range. This is not the case for the fermionic sector. The left panel shows the results in the hadronic sector. At the freeze-out boundary, the values of $\kappa\, \sigma^4/\langle N\rangle$ at various $\mu$-values are given in Fig. \ref{fig:kS4nsT3}. At large $\mu$, both bosonic and fermionic ratios  vanish. With decreasing $\mu$, there is a bounce. It reaching its maximum over $\sim0.25~$GeV. Then both curves come closer to each others. They cross again over $\sim0.2~$GeV. Continue decreasing $\mu$ results in another bounce off. At smaller $\mu$, the fermionic sector switches again to positive values, while the bosonic sector results in  large negative values of $\kappa\, \sigma^4/\langle N\rangle$.  For hadrons, $\kappa\, \sigma^4/\langle N\rangle$ remains almost negative. At very low $\mu$, it turns out to small positive values.

\section{Chemical freeze-out}
\label{sec:chemFO}

In rest frame of produced particle, the hadronic matter can be determined by constant degrees of freedom, for instance, $s/T^3 (4/\pi^2)=const$ \cite{st33}. The chemical freeze-out is related to the particle creation. Therefore, the abundances of different particle species are controlled by chemical potential, which obviously depends on $T$. With the beam energy, $T$ is increasing, while the baryon densities at mid-rapidity is decreasing. The estimation of the macroscopic parameters of the chemical freeze-out can be extracted from particle ratio. These parameters collected over the last two decades seem to fellow regular patterns as the beam energy increases \cite{jean2006,Tawfik:2005qn}. The higher order moments have been suggested to control the chemical freeze-out, so that several conditions have been proposed \cite{HM_FO}.

\begin{figure}[htb]
\includegraphics[angle=-90,width=8.cm]{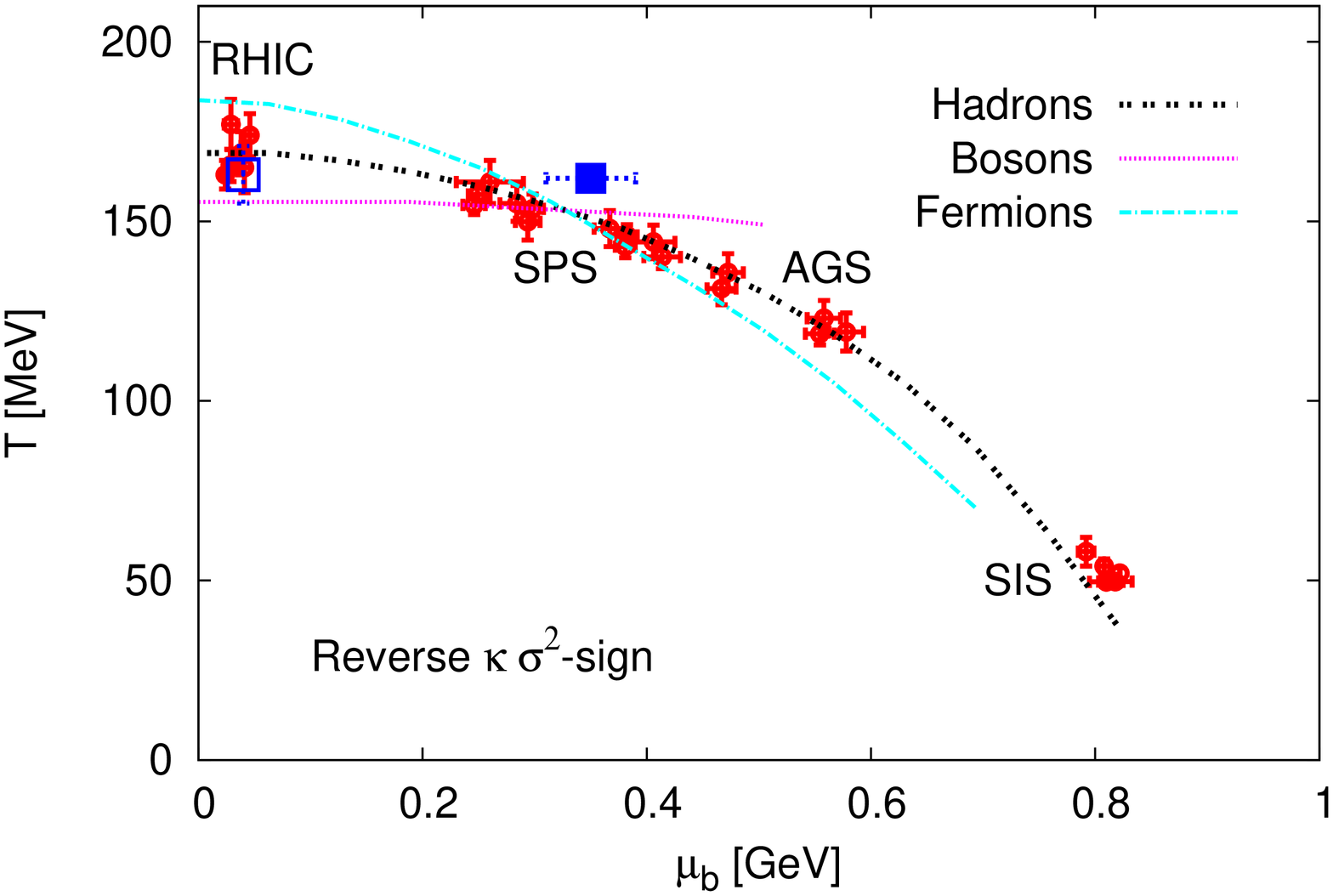}
\includegraphics[angle=-90,width=8.cm]{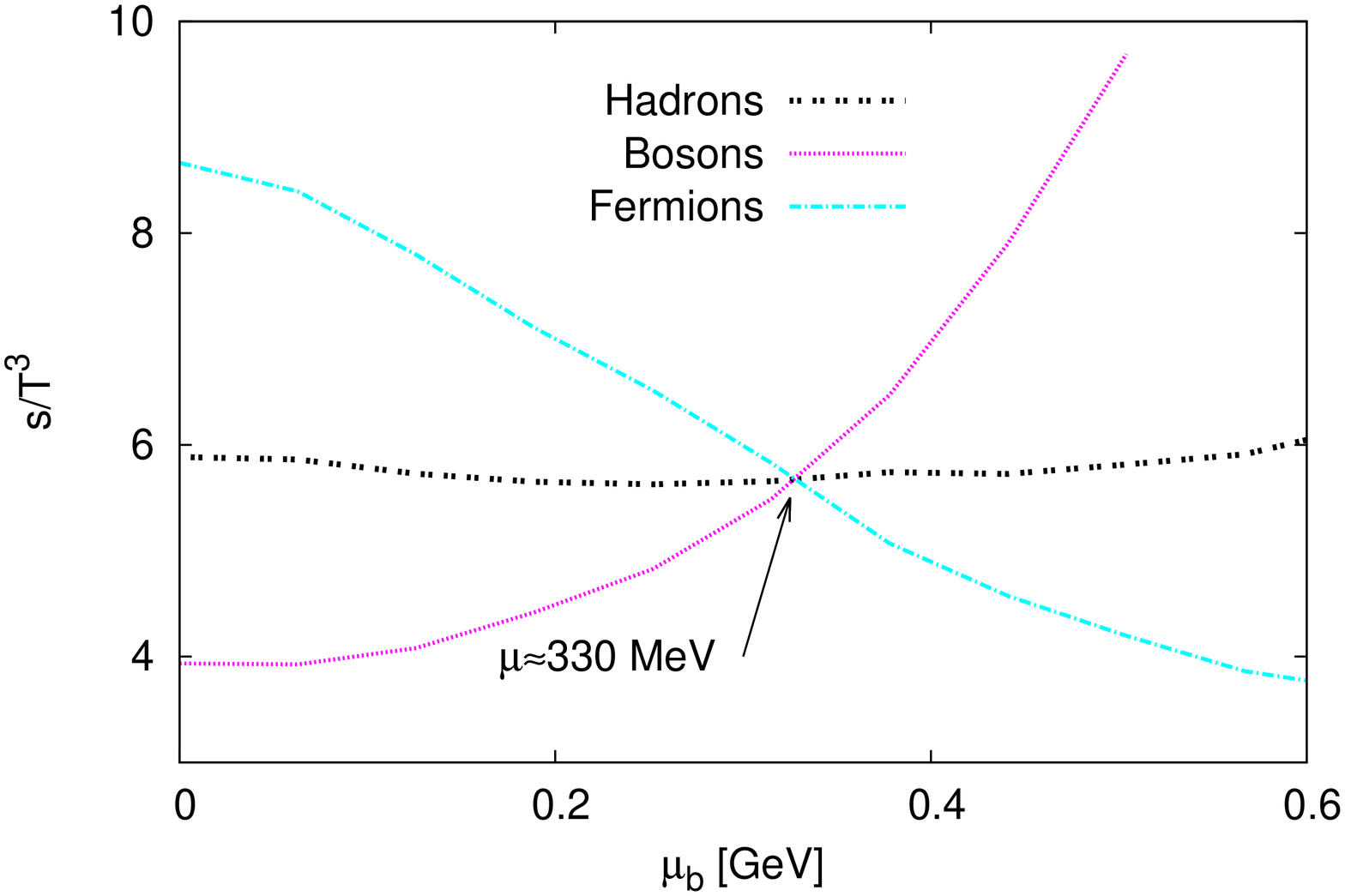}
\caption{Left panel illustrates the chemical freeze-out boundary. The experimental data are given by the solid circles. The curves represent the results of hadron, boson and fermion gas  determined when the sign of ${\kappa}\,\sigma^2$ is flipped. Right panel gives the corresponding $s/T^3$ values. The lattice QCD endpoint is marked by solid square.}
\label{fig:fezeout-sT3} 
\end{figure}

As introduced in section \ref{sec:rmi}, the thermal evolution of ${\kappa}\,\sigma^2$ declines. This result seems to update previous ones \cite{lqcd1a,r42}, where ${\kappa}\,\sigma^2$ is assumed to remain finite and positive with increasing $\mu$. In the present work, we find that the sign of ${\kappa}\,\sigma^2$ is flipped at high $T$ \cite{HM_FO}. It has been found that the $T$- and $\mu$-parameters, at which the sign is flipped are amazingly coincide with the ones of the chemical freeze-out, Fig. \ref{fig:fezeout-sT3}. Also, we find that the freeze-out boundaries of bosons and fermions are crossing at one point located at the hadronic curve. This point is close to the one that lattice QCD calculations suggest for QCD CEP, section \ref{sec:cep}. The results are given in Fig. \ref{fig:fezeout-sT3}.
Vanishing $\kappa\, \sigma^2$ leads to
\bea 
\int_0^{\infty}  \left\{\text{cosh}\left[\frac{\varepsilon_i
      -\mu_i }{T}\right] + 2\right\}\; \text{csch}\left[\frac{\varepsilon_i
      -\mu_i }{2\, T}\right]^4 \; k^2\,dk 
&=& \frac{3\, g_i}{\pi^2}\, \frac{1}{T^3}\, \left[\int_0^{\infty} \left(1-\text{cosh}
  \left[\frac{\varepsilon_i -\mu_i }{T}\right]\right)^{-1}\; k^2\,dk\right]^2, \hspace*{7mm}\label{eq:lsigma2g} \\
\int_0^{\infty}  \left\{\text{cosh}\left[\frac{\varepsilon_i
      -\mu_i }{T}\right] - 2\right\}\; \text{sech}\left[\frac{\varepsilon_i
      -\mu_i }{2\, T}\right]^4 \; k^2\,dk 
&=& \frac{3\, g_i}{\pi^2}\, \frac{1}{T^3}\, \left[\int_0^{\infty} \left(\text{cosh}
  \left[\frac{\varepsilon_i -\mu_i }{T}\right]+1\right)^{-1}\; k^2\,dk\right]^2. \hspace*{10mm}\label{eq:lsigma2fc}
\eea
The rhs and lhs in both expressions can be re-written as 
\bea
16  \frac{\pi^2}{g_i} T^3\, \kappa &=& 48 \frac{\pi^2}{g_i} T^3\, \chi^2,
\eea
which is valid for bosons and fermions. Then, the chemical freeze-out is defined, if the condition
\bea
\kappa(T,\mu) &=& 3\, \chi^2(T,\mu),
\eea 
is fulfilled. At chemical freeze-out curve, a naive estimation leads to $\xi\sim 3^{1/3}~$fm. In doing this, it is assumed that the proportionality coefficients of  $\kappa \sim \xi^7$ and $\chi\sim\xi^2$, are equal.  In the heavy-ion collisions, $\xi$ has been measured \cite{xxii2}. Near a critical point, the experimental value $\sim 2-3\,$fm (only factor 3 larger) agrees well with our estimation.

At the chemical freeze-out curve, the intensive parameters $T$ and $\mu$ which are related to the extensive properties entropy and particle number, respectively, have to be determined over a wide range of beam energies. Left panel of Fig. \ref{fig:fezeout-sT3} collects a large experimental data set. For a recent review, we refer to \cite{jean2006} and the references therein. The double-dotted curve represents a set of $T$ and $\mu$, at which ${\kappa}\,\sigma^2$ vanishes. It is obvious that this curve reproduces very well the experimental data. As given above, at this curve the normalized fourth order moment $\kappa$ is equal to three times the squared susceptibility $\chi$.  This new condition seems to guarantee the condition introduced in \cite{Tawfik:2005qn}; $s/T^3=const.$, right panel of Fig. \ref{fig:fezeout-sT3}, at least over the range $0<\mu<0.6~$GeV. When excluding all degrees of freedom but the fermionic ones, $s/T^3$ decreases with increasing $\mu$ (dashed curve). An opposite dependence is accompanied with the bosonic degrees of freedom (dotted curve).

\section{QCD critical endpoint}
\label{sec:cep}

\begin{figure}[htb]
\includegraphics[angle=-90,width=8.cm]{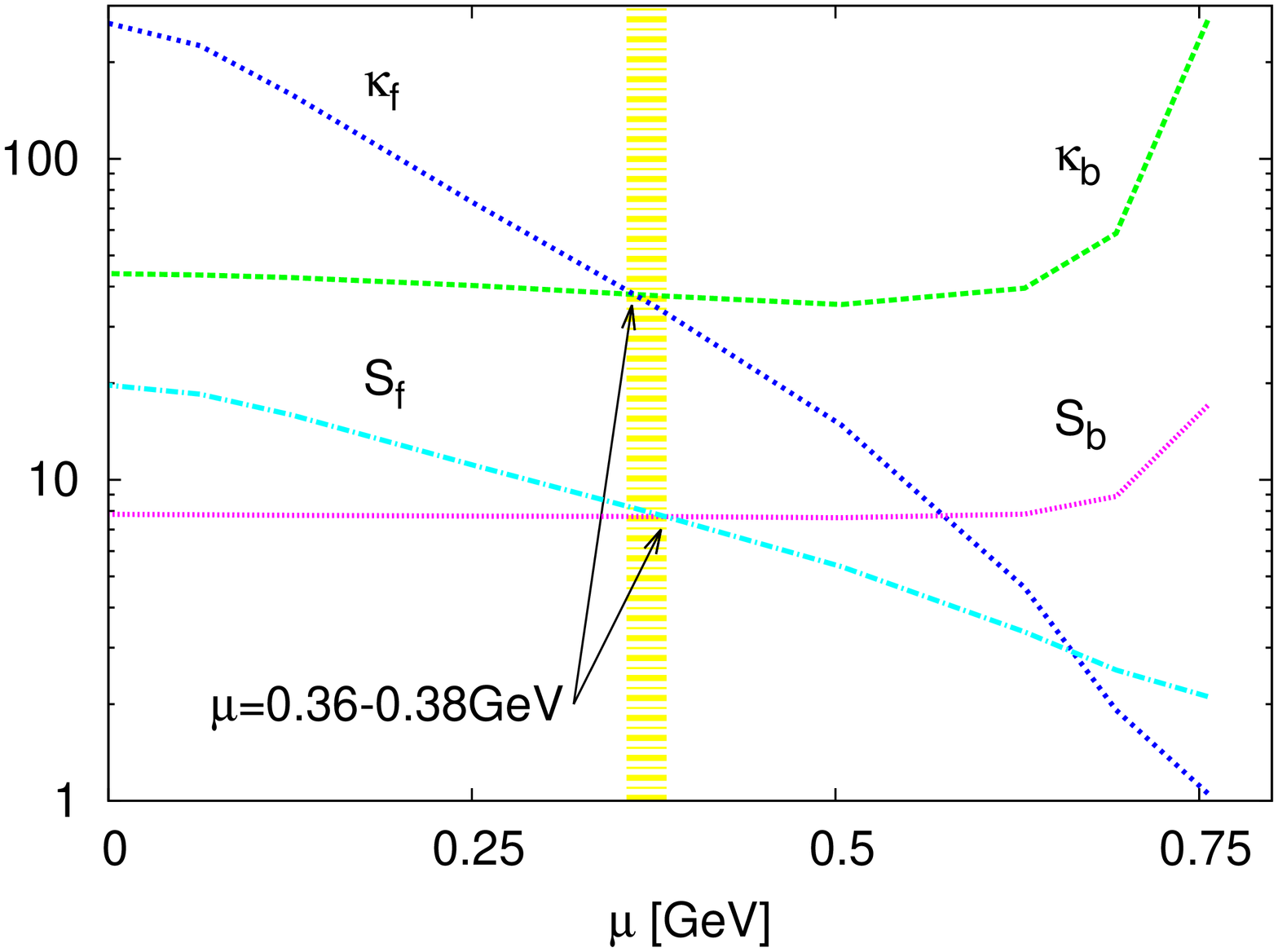}
\includegraphics[angle=-90,width=8.cm]{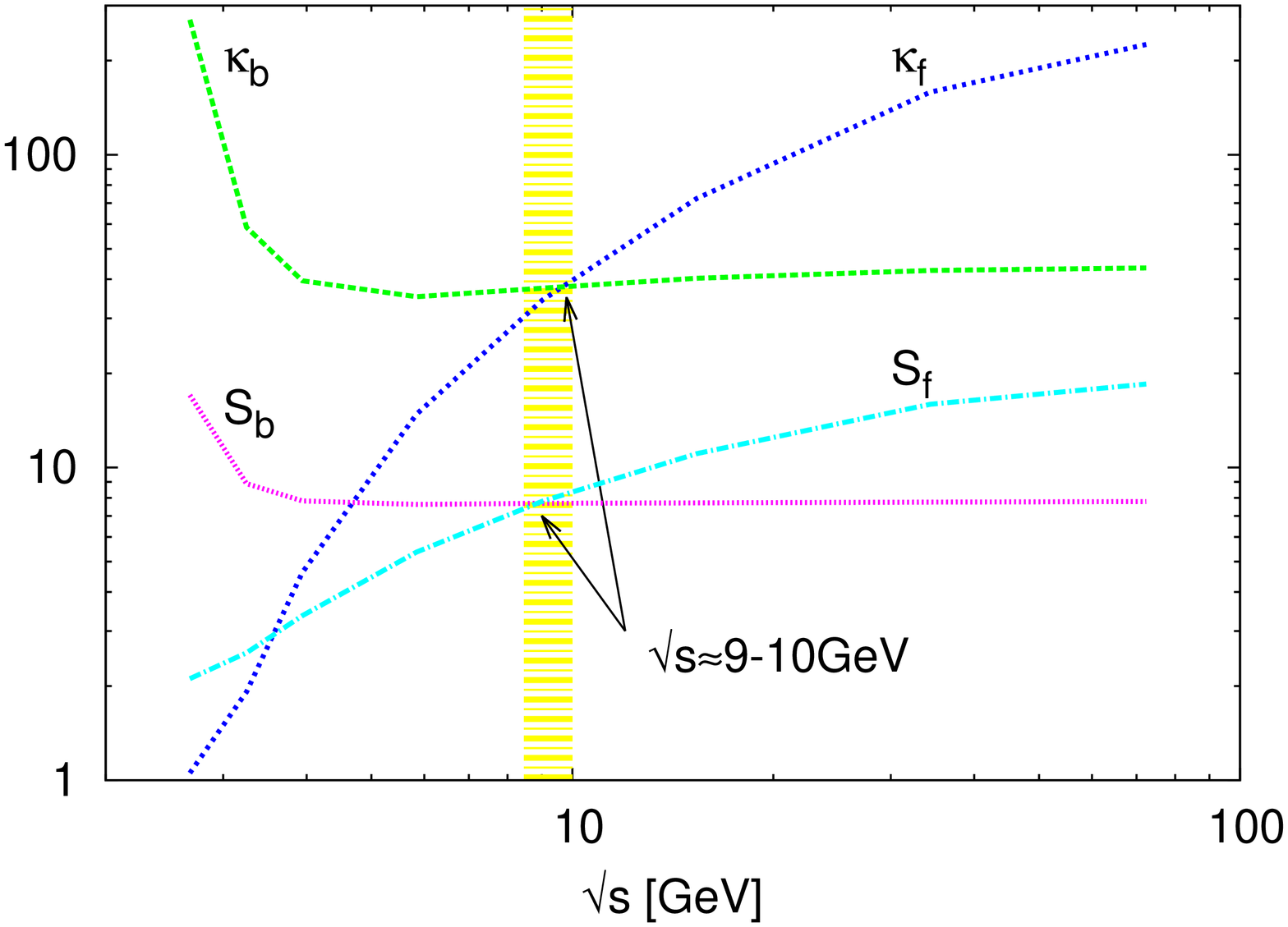}
\caption{Left panel: Skewness and kurtosis at the freeze-out boundary for fermions and bosons. The curves of both quantities are crossing at almost the same $\mu$-value. The corresponding $T$-values are almost equal. }
\label{fig:kSmuu} 
\end{figure}

In right panel of Fig. \ref{fig:fezeout-sT3}, it is interesting to notice that both fermionic and bosonic curves intersect at the hadronic curve at one point. Also, in the freeze-out diagram, all  curves crossing at one point, left panel of Fig. \ref{fig:fezeout-sT3}. Furthermore, based on non-perturbative convergence radius, the critical endpoint as calculated in lattice QCD \cite{endp3b,endp3c} given by solid square in Fig. \ref{fig:fezeout-sT3} is located very near to the crossing point. It is clear that the HRG model does not contain any information on criticality related with the chiral dynamics and singularity in physical observables required to locate the CP. The HRG partition function is an approximation to a non-singular part of the free energy of QCD in the
hadronic phase. It can be used only as the reference for LGT calculations or HIC to verify the critical behavior, but it can not be used as an origin to search for the chiral critical structure in QCD medium. This is the motivation of Fig. \ref{fig:kSmuu}.

In Fig.  \ref{fig:kSmuu}, the skewness and kurtosis of bosons and fermions calculated at the freeze-out curve are given in dependence on $\mu$ (left panel) and center-of-mass energy $\sqrt{s}$ (right panel). It seems that the evolution of fermionic and bosonic skewness and kurtosis coincide at one point, marked with the vertical band. This fact would reflect the nature of the phase transition. It would be the critical endpoint connecting cross-over with the first order deconfinement phase transitions. At QCD CEP, the phase transition is conjectured to be of second order. It is worthwhile to mention the crossing point is amazingly coincident with the QCD CEP measured in lattice QCD \cite{endp3b,endp3c}, Fig. \ref{fig:fezeout-sT3}, regardless its uncertainties.

From Eqs. (\ref{eq:Ssb})-(\ref{eq:Ssf}) and (\ref{eq:Kkb})-(\ref{eq:Kkf}), following expressions have to be solved in $\mu$, individually and/or dependently, in order to determine $\mu$ of the crossing point: 
\bea 
-\, \frac{\int_0^{\infty} \,  
\text{csch}\left[\frac{\varepsilon_i - \mu_i }{2\, T}\right]^4\, \text{sinh} \left[\frac{\varepsilon_i - \mu_i }{T}\right] \; k^2 \, dk}
{\left[\int_0^{\infty} \left(\text{cosh} \left[\frac{\varepsilon_i-\mu_i }{T}\right]-1\right)^{-1}\; k^2 \, dk \right]^{3/2}} 
&=& 16\, \frac{\int_0^{\infty} \,  
\text{csch}\left[\frac{\varepsilon_i - \mu_i }{T}\right]^3\, \text{sinh}
\left[\frac{\varepsilon_i - \mu_i }{2\, T}\right]^4 \; k^2 \, dk}
{\left[\int_0^{\infty} \left(\text{cosh} \left[\frac{\varepsilon_i-\mu_i }{T}\right]+1\right)^{-1}\; k^2 \, dk \right]^{3/2} }, \label{eq:Ssf2} \\
-\frac{\int_0^{\infty}
  \left\{\text{cosh}\left[\frac{\varepsilon_i - \mu_i }{T}\right] + 2\right\} 
\text{csch}\left[\frac{\varepsilon_i - \mu_i }{2\, T}\right]^4 \; k^2 \, dk}
{\left[\int_0^{\infty} \left(1-\text{cosh}\left[\frac{\varepsilon_i-\mu_i }{T}\right]\right)^{-1}\; k^2 \, dk\right]^{2} }  
&=& \frac{\int_0^{\infty}
  \left\{\text{cosh}\left[\frac{\varepsilon_i - \mu_i }{T}\right] - 2\right\} 
\text{sech}\left[\frac{\varepsilon_i - \mu_i }{2\, T}\right]^4 \; k^2 \, dk}
{\left[\int_0^{\infty} \left(\text{cosh}\left[\frac{\varepsilon_i-\mu_i }{T}\right]+1\right)^{-1}\; k^2 \, dk\right]^{2} }.  \hspace*{10mm}\label{eq:Kkf2}
\eea
Due to the mathematical difficulties in dealing with these expressions, the integral over phase space has to be simplified. A suitable simplification is given in section \ref{sec:bessel}. 
\bea
\frac{p(T,\mu_q,\mu_s)}{T^4} &=& \pm \frac{1}{2\pi^2\,T^3}  \sum_{i=1}^{\infty}\,g_i\, m_i^2  \sum_{n=1}^{\infty}\, \frac{(\pm)^{n+1}}{n^2}\, \text{K}_2\left(n\frac{m_i}{T}\right)\, \exp\left[n\frac{(3 n_b+n_s)\mu_q-n_s\mu_s}{T}\right],
\eea 
where $n_b$ and $n_s$ being baryon (strange) quantum number and $\mu_q$ ($\mu_s$) is the baryon  (strange) chemical potential of light and strange quarks, respectively. The quarks chemistry is introduced in section \ref{sec:model}. Accordingly, the difference between baryons and fermions is originated in the exponential function. For simplicity, we consider one fermion and one boson particle. Then, the baryon chemical potential $\mu$  at the chemical freeze-out curve, at which the fermionic and bosonic skewness (or kurtosis) curves of these two particles cross with each other can be given as
\bea
\mu_b = 3 n_b \mu_q &=& T \ln\left[\frac{g_b\, m_b^2\, \text{K}_2\left(\frac{m_b}{T}\right)}{g_f\, m_f^2\, \text{K}_2\left(\frac{m_f}{T}\right)}\right]. \label{eq:mubmusb}
\eea 
In the relativistic limit, $\text{K}_2(m/T)\approx 2 T^2/m^2-1/2$ while in the non-relativistic limit $\text{K}_2(m/T)\approx \sqrt{\pi T/2 m} \exp(-m/T)(1+15 T/8 m)$. It is obvious that the bosonic and fermionic degrees of freedom play an essential role in determining Eq. (\ref{eq:mubmusb}). Furthermore, it seems that the chemical potential of strange quark has no effect at the crossing point.

\begin{figure}[htb]
\includegraphics[angle=-90,width=8.cm]{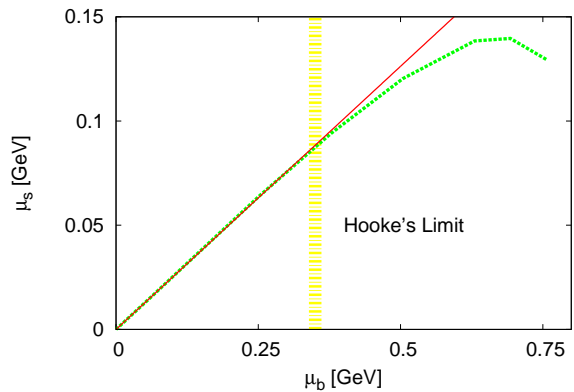}
\caption{$\mu_S$ as a function of $\mu_b$ in hadron resonance gas model (dotted curve). The linear fitting is given by solid curve.}
\label{fig:mub_mus} 
\end{figure}

The dependence of $\mu_S$ on $\mu_b$ as calculated in HRG is given in Fig. \ref{fig:mub_mus}. As mentioned above, $\mu_S$ is calculated to guarantee strange number conservation in heavy-ion collisions. At small $\mu_b$, $\mu_S$ has a linear dependence, $\mu_S=0.25\, \mu_b$ (Hooke's limit). At large $\mu_b$, the dependence  is no longer linear. 

As discussed earlier, the critical behavior and the existence of QCD CEP can be identified by means of signatures sensitive to singular parts of the free energy, especially the ones reflecting dynamical fluctuations of conserved charges, such as baryon number and charge density \cite{cepC}. The reason that the fermionic and bosonic higher order moments are crossing on a point very near the QCD CEP shall be elaborated in a forthcoming work \cite{Tawfik:2013dba}.

\section{Conclusion}

In the present work, the first six non-normalized order moments of the particle multiplicity are calculated in the hadron resonance gas model. A general expressions for arbitrary higher order moments are deduced in terms of lower ones. We concluded that going from lower to higher order moments is possible through adding up a series consists of alls lower order moments plus correlation functions. We studied the thermal evolution of the first four normalized order moments and their products (ratios) at different chemical potentials $\mu$. By doing that, it was possible to evaluate first four normalized moments at the chemical freeze-out curve. The freeze-out curve is characterized by constant $s/T^3$ at all values of $\mu$, where $s$ and $T$ are entropy density and temperature, respectively. It has been found  that non-monotonic behavior reflecting dynamical fluctuation and strong correlations appears starting from the normalized third order moment (skewness $S$). 
Furthermore, non-monotonicity is observed in the normalized fourth order moment, the kurtosis $\kappa$, and its products. These are novel observations. Although HRG is exclusively applicable below $T_c$, i.e. it does not include deconfinement phase transition, it is apparent that the higher order moments are able to give signatures for the critical phase transition. The accuracy of our calculations made it possible to have clear evidences on the phase transition. Based on these findings, we introduced novel conditions characterizing the chemical freeze-out curve and the location of the QCD critical endpoint as follows. The chemical freeze-out curve is described by $\kappa=3\,\chi^2$, where $\chi$ is the susceptibility in particle number, i.e. the second order moment. The location  of QCD critical endpoint (at $T$- and $\mu$-axis) is positioned when the condition $S_b=S_f$ or $\kappa_b=\kappa_f$ is fulfilled. The subscripts $b$ and $f$ refer to bosons and fermions, respectively. Accordingly, the QCD endpoint is positioned at $\mu\sim 350~$MeV and $T\sim162~$MeV. We are able to estimate these two quantities, although the hadron resonance gas model basically does not contain information on criticality related with the chiral dynamics and singularity in physical observables required to locate the critical endpoint. After submitting this work, a new paper posted in arXiv, in which the authors used the second order moment (susceptibility) and the fourth order one (kurtosis) are apparently sensitive to the phase transition \cite{nakamura1}.

\section*{Acknowledgements}

This work is partly supported by the German--Egyptian Scientific Projects (GESP ID: 1378).



\end{document}